\documentclass[twocolumn,aps,superscriptaddress]{revtex4}
\usepackage{amssymb}
\usepackage{amsfonts}
\usepackage{amsmath}
\usepackage{graphicx}
\usepackage{epsf}

\begin{document}

\title{Cyclotron resonance of the magnetic ratchet effect and second harmonic generation in bilayer graphene}

\author{Narjes Kheirabadi}
\affiliation{Physics Department, Lancaster University, Lancaster, LA1 4YB, UK}

\author{Edward McCann}
\affiliation{Physics Department, Lancaster University, Lancaster, LA1 4YB, UK}

\author{Vladimir~I.~Fal'ko}
\affiliation{National Graphene Institute, The University of Manchester, Manchester, M13 9PL, UK}

\begin{abstract}
We model the magnetic ratchet effect in bilayer graphene in which a dc electric current is produced by an
ac electric field of frequency $\omega$ in the presence of a steady in-plane magnetic field and inversion-symmetry breaking.
In bilayer graphene, the ratchet effect is tunable by an external metallic gate
which breaks inversion symmetry.
For zero in-plane magnetic field, we show that trigonal warping and inversion-symmetry breaking
are able to produce a large dc valley current, but not a non-zero total dc charge current.
For the magnetic ratchet in a tilted magnetic field,
the perpendicular field component induces cyclotron motion
with frequency $\omega_c$ and we find that the dc current displays cyclotron resonance at $\omega_c = \omega$,
although this peak in the current is actually smaller than its value at $\omega_c = 0$.
Second harmonic generation, however, is greatly enhanced by resonances at $\omega_c = \omega$
and $\omega_c = 2\omega$ for which the current is generally much larger than at $\omega_c = 0$.
\end{abstract}

\maketitle

\section{Introduction}

\begin{figure}[t]
   \centering
   \includegraphics[scale=0.44]{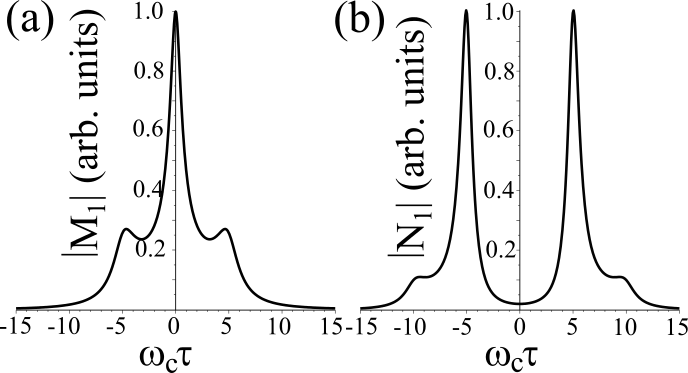}
   \caption{(a) Magnitude of the non-linear coefficient $|M_1|$, Eq.~(\ref{m1approx}), describing the
   magnetic ratchet effect as a function of cyclotron frequency $\omega_c$ for fixed $\omega$.
   (b) Magnitude of the non-linear coefficient $|N_1|$, Eq.~(\ref{n1approx}), describing
   second harmonic generation as a function of $\omega_c$.
   For both plots, $\omega \tau = 5$ and $\tau_2 = \tau$.
   }
    \label{fig:cycl}
\end{figure}

Recently, there has been interest in the magnetic ratchet effect in
two-dimensional electron systems such as semiconductor quantum wells~\cite{falko89,tara08,tara11,gan14,budkin14,bud16},
graphene~\cite{nalitov12,drex13,olbr16} and bilayer graphene~\cite{khe16}.
It is a non-linear effect~\cite{mik07,glazgan} producing a dc electric current
in response to ac laser light in the presence of a steady in-plane magnetic
field and broken inversion symmetry.
Here, we consider the magnetic ratchet in bilayer graphene~\cite{novo06,mccfal06,mccab07,mcckos13,roz16,khe16}
for which inversion asymmetry is tunable by applying
a gate voltage~\cite{mccfal06,mcc06,wu12,brun15}, and we take into account a tilted magnetic field.
If the magnetic field has a perpendicular component, the efficiency of the ratchet
should be dramatically increased under cyclotron resonance conditions~\cite{bud16}
when the cyclotron frequency $\omega_c$ is close to the ac field frequency $\omega$.
For the dc current, we find that there is indeed a resonance at $\omega_c = \omega$, but,
as a function of $\omega_c$, the dc current is actually larger at $\omega_c = 0$, Fig.~\ref{fig:cycl}(a).
For second harmonic generation~\cite{dean09,dean10,gla11,mik11,byk12,wu12,gol14,lin14,brun15}, however,
we find there are resonances at $\omega_c = \omega$
and $\omega_c = 2\omega$ which are generally much stronger than the current at $\omega_c = 0$, Fig.~\ref{fig:cycl}(b).

We begin by summarizing our results. In bilayer graphene, the intraband contribution
(which is relevant in the semiclassical regime $\epsilon_F \gg \hbar\omega$ where $\epsilon_F$ is the Fermi level)
to the dc current density $\mathbf{J}^{(0)}$ is given by
\begin{eqnarray}
\mathbf{J}^{(0)} &\approx& \Big\{ \left[ \left( \mathbf{B}_{\parallel} \times \mathbf{\hat{n}}_z \right) \cdot \mathbf{E} \right] \mathbf{E}^{\ast}
+ \left[ \left( \mathbf{B}_{\parallel} \times \mathbf{\hat{n}}_z \right) \cdot \mathbf{E}^{\ast} \right] \mathbf{E} \nonumber \\
&& \qquad \qquad
- \left( \mathbf{B}_{\parallel} \times \mathbf{\hat{n}}_z \right) \left| \mathbf{E} \right|^2 \Big\} \mathrm{Re} \left( M_1 \right) \nonumber \\
&& + \Big\{ \mathbf{E} \left( \mathbf{B}_{\parallel} \cdot \mathbf{E}^{\ast} \right) +
\mathbf{E}^{\ast} \left( \mathbf{B}_{\parallel} \cdot \mathbf{E} \right) \nonumber \\
&& \qquad \qquad
- \mathbf{B}_{\parallel} \left| \mathbf{E} \right|^2 \Big\} \mathrm{Im} \left( M_1 \right) , \label{j0intro}
\end{eqnarray}
for normally-incident radiation with an in-plane alternating electric field $\mathbf{E}(t)$
of angular frequency $\omega$,
\begin{eqnarray}
\mathbf{E}(t) = \mathbf{E} e^{-i\omega t} + \mathbf{E}^{\ast} e^{i\omega t} , \label{ace}
\end{eqnarray}
and in-plane magnetic field $\mathbf{B}_{\parallel}$,
$\mathbf{\hat{n}}_z$ is a unit vector in the perpendicular ($z$) direction.
We include terms that are second order in electric field (i.e. quadratic in electric field amplitudes $\mathbf{E}$ and $\mathbf{E}^{\ast}$), and linear in $\mathbf{B}_{\parallel}$.
Second order (in electric field) effects generally require breaking of spatial inversion symmetry~\cite{ped96,shen03,boyd08}, and, here, the coefficient $M_1$
changes sign upon $z \rightarrow - z$ inversion.

Eq.~(\ref{j0intro}) describes the contribution arising from a
perfectly quadratic dispersion relation $\epsilon = p^2/2m$ ($p$ is the magnitude of momentum
and $m$ is mass) when the relaxation rates are independent of energy,
and this contribution will generally dominate in bilayer graphene
(there will be small corrections
when these conditions are not exactly met as described in detail in Section~\ref{s:be}).
Parameter $M_1$ describes the response to incoming linearly polarized light
and we find that
\begin{eqnarray}
M_1 \approx - \frac{g e^3 p^2}{16 \pi^2 \hbar^4}
\frac{\left(2\Lambda_1 + \epsilon \Lambda_1^{\prime}\right)}{\Upsilon^{0,1}\Upsilon^{0,2}}
\left( \frac{1}{\Upsilon^{1,1}} + \frac{1}{\Upsilon^{-1,1}}\right) , \label{m1approx}
\end{eqnarray}
where $g$ is a degeneracy factor ($g=4$ for spin and valley in graphene)
and the electronic charge is $-e$, $e>0$.
For a particular material, parameter $\Lambda_1$
(which is independent of $\omega$ and $\omega_c$)
characterizes the strength of scattering in
the presence of an in-plane magnetic field and $z \rightarrow - z$ asymmetry (we will
present the explicit form of $\Lambda_1$ for bilayer graphene later).
The functions $\Upsilon^{\ell,j}$, with integer $\ell$ and $j$,
where
\begin{eqnarray}
\Upsilon^{\ell,j} = \tau_{|j|}^{-1} - i \ell \omega + i j \omega_c \, ;
\qquad \tau_{1} \equiv \tau \, ,
\end{eqnarray}
account for the dependence of the current on electric field frequency $\omega$
and cyclotron frequency $\omega_c = e B_{\perp} v_g / p$ including the cyclotron
resonance effect \cite{bud16,olb13,dan15}.
Here $B_{\perp} = |\mathbf{B}_{\perp}|$ and we consider
$\mathbf{B}_{\perp} = B_{\perp} \mathbf{\hat{n}}_z$.
Also, $v_g$ is the group velocity $v_g = d\epsilon/dp$, so
$\omega_c = e B_{\perp} v^2/\epsilon$ \cite{and06} for linear dispersion $\epsilon = vp$,
and $\omega_c = e B_{\perp} /m$ \cite{hei10} for quadratic dispersion $\epsilon = p^2/(2m)$.
Parameters $\tau_{1}$ and $\tau_{2}$ are relaxation times for the first and second
angular harmonics of the electronic distribution, respectively
($\tau_{1}$ is the usual momentum relaxation rate that we denote as $\tau$ from now on).

Figure~\ref{fig:cycl}(a) shows the magnitude of the non-linear coefficient $|M_1|$
as a function of cyclotron frequency $\omega_c$ for fixed $\omega$.
Although there is a noticeable resonance for $\omega_c = \omega$
[due to the presence of $\Upsilon^{\pm 1,1}$ in Eq.~(\ref{m1approx}], the ratchet
effect is actually strongest for $\omega_c = 0$ because of the product
$\Upsilon^{0,1}\Upsilon^{0,2}$.
This product arises because of the need to couple with dc components
of the electronic distribution in order to create dc current.

The second harmonic current density $\mathbf{J}^{(2)}$ is given by
\begin{eqnarray}
\mathbf{J}^{(2)} &\approx&
2 \mathrm{Re} \Big\{ \Big[ 2 \left[ \left( \mathbf{B}_{\parallel} \times \mathbf{\hat{n}}_z \right) \cdot \mathbf{E} \right] \mathbf{E} \nonumber \\
&& \qquad \qquad
- \left( \mathbf{B}_{\parallel} \times \mathbf{\hat{n}}_z \right) \mathbf{E}^2 \Big] N_1 e^{-2i \omega t} \Big\} \nonumber \\
&& - 2 \mathrm{Re} \Big\{ \left[ 2 \left( \mathbf{B}_{\parallel} \cdot \mathbf{E} \right) \mathbf{E}
- \mathbf{B}_{\parallel} \mathbf{E}^2 \right] N_2 e^{-2i \omega t} \Big\} , \label{j2intro}
\end{eqnarray}
where $\mathbf{E}^2 \equiv \mathbf{E} \cdot \mathbf{E} = E_x^2 + E_y^2$.
This describes the contribution arising from a
perfectly quadratic dispersion relation when the relaxation rates are independent of energy
(there will be small corrections
when these conditions are not exactly met as described in detail in Section~\ref{s:be}).
These terms describe an effect similar to the Faraday effect~\cite{ped96} in that incoming
plane-polarized light results in the emission of a plane-polarized second harmonic
with the magnetic field contributing to a rotation of the angle of polarization.
Incoming circularly-polarized light results in circularly-polarized second harmonic
generation.
The coefficients $N_1$, $N_2$ are given by
\begin{eqnarray}
N_1 &=& - \frac{g e^3 p^2}{32 \pi^2 \hbar^4}
\left(\Lambda_1 + \epsilon \Lambda_1^{\prime}\right) \nonumber \\
&& \quad \times \!
\left( \frac{1}{\Upsilon^{1,1}\Upsilon^{2,2}\Upsilon^{2,1}}
+ \frac{1}{\Upsilon^{1,-1}\Upsilon^{2,-2}\Upsilon^{2,-1}} \right) \! , \label{n1approx} \\
N_2 &=& - \frac{i g e^3 p^2}{32 \pi^2 \hbar^4}
\left(\Lambda_1 + \epsilon \Lambda_1^{\prime}\right) \nonumber \\
&& \quad \times \!
\left( \frac{1}{\Upsilon^{1,1}\Upsilon^{2,2}\Upsilon^{2,1}}
- \frac{1}{\Upsilon^{1,-1}\Upsilon^{2,-2}\Upsilon^{2,-1}} \right) \! . \label{n2approx}
\end{eqnarray}
Figure~\ref{fig:cycl}(b) shows the magnitude of the non-linear coefficient $|N_1|$
as a function of cyclotron frequency $\omega_c$ for fixed $\omega$
(note that $|N_2|$ shows almost the same dependence on $\omega_c$,
both qualitatively and quantitatively, except that $|N_2|=0$ for $\omega_c = 0$).
In stark contrast to $|M_1|$, the resonance at $\omega_c = \omega$
is far stronger than the signal at $\omega_c = 0$ because
there are no dc components of the electronic distribution
involved ({\em i.e.} no $\Upsilon^{0,j}$ factors), and there is also a resonance
at $\omega_c = 2 \omega$.

The combinations of electric and magnetic fields in Eqs.~(\ref{j0intro},\ref{j2intro}) satisfy spatial symmetries
including rotations in the two-dimensional plane ($x$-$y$) of the sample.
As $\mathbf{B}_{\parallel}$ is an axial vector, the combination $\mathbf{B}_{\parallel} \times \mathbf{\hat{n}}_z$,
appearing in the $\mathrm{Re} \left( M_1 \right)$ and $N_1$ terms, behaves as a true vector.
However, factors in the $\mathrm{Im} \left( M_1 \right)$ and $N_2$ terms containing $\mathbf{B}_{\parallel}$ appear to break some reflection symmetries
({\em e.g.} reflection in a plane perpendicular to the sample, such as the $y$-$z$ plane)
but they are in fact satisfied because $\mathrm{Im} \left( M_1 \right)$ and $N_2$ are odd functions
of the combination $\mathbf{\hat{n}}_z \cdot \mathbf{B}_{\perp}$ and, thus, change sign upon
such reflections.
Hence, $\mathrm{Im} \left( M_1 \right) = 0$ and $N_2=0$ when
$\mathbf{B}_{\perp} = 0$ whereas $\mathrm{Re} \left( M_1 \right)$ and $N_1$ are even functions of
$\mathbf{\hat{n}}_z \cdot \mathbf{B}_{\perp}$ and are generally non-zero for $\mathbf{B}_{\perp} = 0$.

In the next Section, we describe a phenomenological Drude model in which the ratchet effect
may be viewed as arising from the motion of classical particles in the presence of friction
created by the in-plane magnetic field. This model correctly accounts
for the combinations of fields in Eqs.~(\ref{j0intro},\ref{j2intro}), but doesn't quite
account for the frequency dependences in Eqs.~(\ref{m1approx},\ref{n1approx},\ref{n2approx}) because it only contains one relaxation rate $\tau$.
However, in Section~\ref{s:be}, we use the Boltzmann equation
to derive general equations for the dc current and second harmonic for a magnetic ratchet in an arbitrary two-dimensional electron system with $z \rightarrow - z$ asymmetry and an isotropic dispersion $\epsilon (p)$,
including cyclotron motion, too.
Section~\ref{bg} describes scattering in the presence of an in-plane magnetic field in bilayer graphene
in order to apply the general equations to it, resulting in the simplified expressions Eqs.~(\ref{j0intro},\ref{j2intro}).
In Section~\ref{s:valley}, we show that, for $\mathbf{B}_{\parallel} = 0$, trigonal warping in bilayer graphene
leads to a valley current of large magnitude which, when summed over both valleys, yields zero total current,
unless there is an additional source of valley polarization.

\section{Phenomenological Drude model}

The expressions~(\ref{j0intro},\ref{j2intro}) for non-linear current densities
may be viewed as being due to the motion of classical particles in the presence of friction
created by the in-plane magnetic field.
Previously \cite{falko89,tara11}, a classical, Drude model has been used to provide a simple picture
of the origin of ratchet current for intraband transitions. Here, we generalize this approach for the case where the non-linearity is produced by the
in-plane magnetic field, and we consider cyclotron motion and second harmonic generation, too. We consider the equation of motion for the average
drift velocity $\mathbf{v}_d$ per electron in a two-dimensional system:
\begin{eqnarray}
m \frac{d \mathbf{v}_d}{dt} = - e\mathbf{E} - e(\mathbf{v}_d \times \mathbf{B}_{\perp})
- \frac{m\mathbf{v}_d}{\tau} + \mathbf{F} (\mathbf{v}_d, \mathbf{B}_{\parallel} ) , \label{drude2}
\end{eqnarray}
where $\tau$ is the momentum relaxation time,
and the perpendicular magnetic field $\mathbf{B}_{\perp}$ and in-plane ac electric field
$\mathbf{E}(t)$, Eq.~(\ref{ace}), enter via the Lorentz force.
The term $\mathbf{F} (\mathbf{v}_d, \mathbf{B}_{\parallel} )$ describes friction
due to the presence of the in-plane magnetic field $\mathbf{B}_{\parallel}$ which
introduces non-linearity into the system. We assume this term is quadratic in velocity $\mathbf{v}_d$
and linear in magnetic field $\mathbf{B}_{\parallel}$:
\begin{eqnarray*}
\mathbf{F} (\mathbf{v}_d, \mathbf{B}_{\parallel} ) = \alpha
\left\{ 2 \left[ ( \mathbf{B}_{\parallel} \times \mathbf{\hat{n}}_z ) \cdot \mathbf{v}_d \right]
\mathbf{v}_d
- ( \mathbf{B}_{\parallel} \times \mathbf{\hat{n}}_z ) |\mathbf{v}_d|^2 \right\} \! ,
\end{eqnarray*}
where $\alpha$ is a phenomenological parameter that characterizes the material properties of the particular system in question.
This form of the friction term is obtained by requiring that it behaves as a true vector
rather than, say, an axial vector (as $\mathbf{B}_{\parallel}$ is an axial vector, the combination $\mathbf{B}_{\parallel} \times \mathbf{\hat{n}}_z$ behaves as a true vector).
In principle, there should be a second phenomenological parameter
to characterize the relative weight of the two terms in $\mathbf{F} (\mathbf{v}_d, \mathbf{B}_{\parallel} )$,
but, for simplicity, we insert its value above [it is determined by demanding that we obtain non-linear
contributions with the correct combinations of fields Eqs.~(\ref{j0intro},\ref{j2intro})].

To solve Eq.~(\ref{drude2}) we use a harmonic expansion of the velocity,
\begin{eqnarray*}
\mathbf{v}_d(t) = \sum_{\ell=0,\pm 1, \ldots} \mathbf{v}_d^{(\ell)} e^{-in\omega t} \, ,
\end{eqnarray*}
which yields coupled equations for the coefficients $\mathbf{v}_d^{(\ell)}$:
\begin{eqnarray}
-i\ell\omega \mathbf{v}_d^{(\ell)} &=&
- \frac{e\mathbf{E}}{m} \delta_{\ell , 1} - \frac{e\mathbf{E}^{\ast}}{m} \delta_{\ell , -1}
- \omega_c (\mathbf{v}_d^{(\ell)} \times \mathbf{\hat{n}}_z)
- \frac{\mathbf{v}_d^{(\ell)}}{\tau} \nonumber \\
&& + \frac{\alpha}{m} \sum_n
\Big\{ 2 \left[ ( \mathbf{B}_{\parallel} \times \mathbf{\hat{n}}_z ) \cdot \mathbf{v}_d^{(n)} \right] \mathbf{v}_d^{(\ell -n)} \nonumber \\
&& \qquad \qquad
- ( \mathbf{B}_{\parallel} \times \mathbf{\hat{n}}_z ) \left[ \mathbf{v}_d^{(n)} \cdot \mathbf{v}_d^{(\ell -n)} \right] \Big\} \, . \label{drudec}
\end{eqnarray}
Neglecting the in-plane magnetic field, the linear harmonics are given by
\begin{eqnarray*}
\left(\mathbf{v}_d^{(1)}\right)_x &=& - \frac{e \tau}{m} \left[\frac{(1-i\omega \tau)E_x - \omega_c \tau E_y}{(1-i\omega \tau)^2 + (\omega_c \tau)^2}\right] \, , \\
\left(\mathbf{v}_d^{(1)}\right)_y &=& - \frac{e \tau}{m} \left[\frac{(1-i\omega \tau)E_y + \omega_c \tau E_x}{(1-i\omega \tau)^2 + (\omega_c \tau)^2}\right] \, ,
\end{eqnarray*}
and $\mathbf{v}_d^{(-1)} = \left( \mathbf{v}_d^{(1)} \right)^{\ast}$.
Since current density is related to drift velocity by $\mathbf{J}(t) = - n e \mathbf{v}_d (t)$
where $n$ is the electron number density~\cite{hei10}, the linear current density may be written
as $\mathbf{J}^{(1)} = - 2 n e \mathrm{Re} \{\mathbf{v}_d^{(1)} e^{-i\omega t}\}$
which yields $\mathbf{J}^{(1)} = 2 \mathrm{Re} \{ \sigma \mathbf{E} e^{-i\omega t}\}$
where the conductivity tensor $\sigma$ has components
\begin{eqnarray}
\sigma_{xx} = \sigma_{yy} &=& \frac{(1-i\omega \tau)\sigma_0}{(1-i\omega \tau)^2 + (\omega_c \tau)^2} , \label{long} \\
\sigma_{xy} = - \sigma_{yx} &=& - \frac{\omega_c \tau \sigma_0}{(1-i\omega \tau)^2 + (\omega_c \tau)^2} , \label{hall}
\end{eqnarray}
with dc Drude conductivity $\sigma_0 = n e^2 \tau / m$, as expected~\cite{hei10}.

The in-plane magnetic field introduces non-linearity which couples non-linear harmonics to the linear one, Eq.~(\ref{drudec}). To first order in $\mathbf{B}_{\parallel}$, we find that the ratchet current density is given by Eq.~(\ref{j0intro})
and the Drude expression for the non-linear coefficient $M_1$ is
\begin{eqnarray*}
M_1^{D} = - \frac{\alpha n e^3}{m^3} \frac{1}{(\Upsilon^{0,1})^2}
\left( \frac{1}{\Upsilon^{1,1}} + \frac{1}{\Upsilon^{-1,1}} \right) \, .
\end{eqnarray*}
Likewise, the second harmonic current density
is given by Eq.~(\ref{j2intro})
and the Drude expressions for the non-linear coefficients $N_1$, $N_2$ are
given by
\begin{eqnarray*}
N_1^{D} &=& - \frac{\alpha n e^3}{2 m^3}
\left[ \frac{1}{(\Upsilon^{1,1})^2 \Upsilon^{2,1}}
+ \frac{1}{(\Upsilon^{1,-1})^2 \Upsilon^{2,-1}} \right] , \\
N_2^{D} &=& - \frac{i \alpha n e^3}{2 m^3}
\left[ \frac{1}{(\Upsilon^{1,1})^2 \Upsilon^{2,1}}
- \frac{1}{(\Upsilon^{1,-1})^2 \Upsilon^{2,-1}} \right] .
\end{eqnarray*}
Comparison with the coefficients derived using the Boltzmann equation
Eqs.~(\ref{m1approx},\ref{n1approx},\ref{n2approx}) shows that although the simple Drude model
correctly produces the correct combinations of fields Eqs.~(\ref{j0intro},\ref{j2intro}), it
doesn't quite account for the frequency dependence because it only contains one relaxation rate $\tau$
[the parameter $\tau_{2}$ describing relaxation
of the second angular harmonic of the electronic distribution is not included in Eq.~(\ref{drude2})].
In order to accurately describe temporal {\em and} spatial relaxation
of the electronic distribution it is necessary to employ the Boltzmann equation,
as described in the next Section.

\section{Boltzmann equation}\label{s:be}

In this section, we derive the intraband contribution to
the second order (in electric field) non-linear conductivity
due to the presence of an in-plane magnetic field $\mathbf{B}_{\parallel}$ for an arbitrary two-dimensional electron
system with $z \rightarrow - z$ asymmetry and an isotropic dispersion $\epsilon (p)$.
We consider linear-in-$\mathbf{B}_{\parallel}$ terms and we take into account
the effect of a perpendicular magnetic field $\mathbf{B}_{\perp}$ which
introduces cyclotron motion with cyclotron frequency $\omega_c = e B_{\perp} v_g / p$.
These semi-classical calculations are valid for finite Fermi energy $\epsilon_F$
with $\epsilon_F \gg \{ \hbar / \tau , \hbar \omega \}$, $\omega_c  \tau \ll 1$,
and we also assume that the electrons are degenerate $\epsilon_F \gg k_BT$.

We consider a spatially homogeneous system with electron motion described
by the Boltzmann equation~\cite{hei10},
\begin{eqnarray}
- e \left( \mathbf{E}_{\parallel} + \mathbf{v}_g \times \mathbf{B}_{\perp} \right).\nabla_p f (\mathbf{p},t)
+ \frac{\partial f (\mathbf{p},t)}{\partial t}
= S\!\!\:\{f\} , \label{boltz1}
\end{eqnarray}
where the electron distribution $f (\mathbf{p},t)$ is a function of
momentum $\mathbf{p}$ and time $t$,
$\mathbf{v}_g = v_g \left(\boldsymbol{\hat{\textbf{\i}}}\cos \phi + \boldsymbol{\hat{\textbf{\j}}}\sin\phi\right)$
and $\phi$ is the polar angle of momentum.
The in-plane alternating field $\mathbf{E}_{\parallel}(t)$, Eq.~(\ref{ace}),
and the perpendicular magnetic field $\mathbf{B}_{\perp} = B_{\perp} \mathbf{\hat{n}}_z$
are accounted for by the Lorentz force, while the in-plane magnetic field $\mathbf{B}_{\parallel}$
enters via the the collision integral $S\!\!\:\{f\}$,
\begin{eqnarray}
S\!\!\:\{f\} = \sum_{\mathbf{p}^{\prime}} \left[ W_{\mathbf{p}\mathbf{p}^{\prime}}
f (\mathbf{p}^{\prime},t) - W_{\mathbf{p}^{\prime}\mathbf{p}}
f (\mathbf{p},t) \right] . \label{colint}
\end{eqnarray}
Here $W_{\mathbf{p}^{\prime}\mathbf{p}}$ is the scattering rate describing
scattering from initial $|\mathbf{p}\rangle$ to final state $|\mathbf{p}^{\prime}\rangle$
in the presence of a scattering potential $\delta H$, as given by Fermi's golden rule,
\begin{eqnarray}
W_{\mathbf{p'p}} = \frac{2\pi}{\hbar}\left|\langle \mathbf{p'}\left| \delta H \right| \mathbf{p} \rangle \right|^2 \delta (\epsilon_\mathbf{p}-\epsilon_\mathbf{p'}) \, . \label{formulsp}
\end{eqnarray}
We consider static impurities
\begin{eqnarray*}
\delta H = \sum_{j=1}^{N_i} \hat{Y} u \! \left( \mathbf{r} - \mathbf{R}_j \right) , \label{static}
\end{eqnarray*}
where $N_i$ is the number of impurities,
$u \! \left( \mathbf{r} - \mathbf{R}_j \right)$ describes the spatial dependence
of the impurity potential.
The dimensionless matrix $\hat{Y}$ takes account of any additional degree of freedom
related to structure within the unit cell, for example $A$/$B$ lattice in graphene.
We neglect interference between different impurities and use the Fourier transform of the impurity potential,
\begin{eqnarray*}
\tilde{u} \! \left( \mathbf{q} \right) = \int d^2r \, u \! \left( \mathbf{r} \right)
e^{-i\mathbf{q}.\mathbf{r}/\hbar}
\, .
\end{eqnarray*}
When evaluating the scattering rate~(\ref{formulsp}), we expand the states
$|\mathbf{p}\rangle$, $|\mathbf{p}^{\prime}\rangle$ in powers of the in-plane magnetic field
$\mathbf{B}_{\parallel}$. For zeroth order, $\mathbf{B}_{\parallel} = 0$, we recover the usual
relaxation rates for the $j$th angular harmonics of the electronic distribution,
\begin{eqnarray}
\tau_{|j|}^{-1} &=& \frac{2\pi}{\hbar} \sum_{\mathbf{p}^{\prime}}
\left| \langle \mathbf{p}^{\prime} | \delta H | \mathbf{p} \rangle \right|^2
\delta \! \left( \epsilon_{\mathbf{p}} - \epsilon_{\mathbf{p}^{\prime}} \right) \nonumber \\
&& \qquad \qquad \qquad \times \left[ 1 - \cos \left( j \left[ \phi^{\prime} - \phi \right] \right) \right] \, , \label{tau}
\end{eqnarray}
and we write $\tau_{1} \equiv \tau$.
To linear order in $\mathbf{B}_{\parallel}$, we find that the
scattering rate may be written generically as
\begin{eqnarray}
&& \delta W_{\mathbf{p}^{\prime}\mathbf{p}} = \frac{1}{L^2} \left| \tilde{u} \! \left( \mathbf{p}^{\prime} - \mathbf{p} \right) \right|^2
\delta \! \left( \epsilon_{\mathbf{p}} - \epsilon_{\mathbf{p}^{\prime}} \right) \label{wgen} \\
&& \!\!\!\!\times \bigg\{ \!\!
\left( \Omega - \Omega_c \cos [2(\phi^{\prime} - \phi)] \right) \!\!
\left[ B_x \left( p_y^{\prime} + p_y \right) - B_y \left( p_x^{\prime} + p_x \right) \right] \nonumber \\
&&
\quad  + \, \Omega_s \sin [2(\phi^{\prime} - \phi)] \!\!
\left[ B_x \left( p_x^{\prime} - p_x \right) + B_y \left( p_y^{\prime} - p_y \right) \right] \!\! \bigg\} \, , \nonumber
\end{eqnarray}
where $\Omega$, $\Omega_c$, $\Omega_s$ are angle-independent factors that depend on
specific material properties.

In order to solve the Boltzmann equation~(\ref{boltz1}), we use
polar coordinates $(p,\phi)$ for momentum and expand the distribution function in terms of $\phi$ and $t$ harmonics with coefficients $f_m^{(n)}$:
\begin{eqnarray}
f (\mathbf{p},t) = \sum_{n,m} f_m^{(n)} e^{im\phi - i n \omega t} , \label{expand}
\end{eqnarray}
where $m$, $n$ are integers. We also perform an harmonic expansion of the impurity potential,
\begin{eqnarray*}
\left| \tilde{u} \! \left( \mathbf{p}^{\prime} - \mathbf{p} \right) \right|^2
= \sum_m \nu_m e^{im (\phi^{\prime} - \phi)} \, ,
\end{eqnarray*}
with the constraint that $\nu_{-m} = \nu_m$ as it is an even function of
$(\phi^{\prime} - \phi)$.
Then, we multiply the Boltzmann equation by a factor $\exp ( - i j \phi + i \ell \omega t )$,
where $j$, $\ell$ are integers, and integrate over a period $2\pi$ of angle $\phi$
and a period of time $t$. This results in coupling between different harmonic coefficients:
\begin{eqnarray}
f_j^{(\ell)} \!\! \left( \tau_{|j|}^{-1} - i \ell \omega + i j \omega_c \right) \!
&=& \! \alpha_{j-1} f_{j-1}^{(\ell - 1)} + \eta_{j+1} f_{j+1}^{(\ell - 1)} \label{coupled} \\
&& \!\!\! \!\!\! \!\!\! \!\!\! \!\!\!
+ \, \tilde{\alpha}_{j-1} f_{j-1}^{(\ell + 1)} + \tilde{\eta}_{j+1} f_{j+1}^{(\ell + 1)} + \delta \!S_{j}^{(\ell)}  . \nonumber
\end{eqnarray}
Operators $\alpha$, $\eta$ are linear in the electric field,
\begin{eqnarray*}
\alpha_{j} &=& \frac{e \left(E_x - i E_y \right)}{2} \left( - \frac{j}{p} + \frac{\partial}{\partial p} \right) \, , \\
\tilde{\alpha}_{j} &=& \frac{e \left(E_x^{\ast} - i E_y^{\ast} \right)}{2} \left( - \frac{j}{p} + \frac{\partial}{\partial p} \right) \, , \\
\eta_{j} &=& \frac{e \left(E_x + i E_y \right)}{2} \left( \frac{j}{p} + \frac{\partial}{\partial p} \right) \, , \\
\tilde{\eta}_{j} &=& \frac{e \left(E_x^{\ast} + i E_y^{\ast} \right)}{2} \left( \frac{j}{p} + \frac{\partial}{\partial p} \right) \, , \\
\end{eqnarray*}
and factors $\delta \!S_{j}^{(\ell)}$ account for the linear-in-$\mathbf{B}_{\parallel}$
correction to scattering, the relevant ones have small values of $j$:
\begin{eqnarray*}
\delta \!S_{0}^{(\ell)} \!&=&\! 0 , \\
\delta \!S_{1}^{(\ell)} \!&=&\! \tfrac{1}{2} p \, \Gamma (\epsilon) \left( B_y - i B_x \right)
\Lambda_1 f_{2}^{(\ell)} , \\
\delta \!S_{-1}^{(\ell)} \!&=&\! \tfrac{1}{2} p \, \Gamma (\epsilon) \left( B_y + i B_x \right)
\Lambda_1 f_{-2}^{(\ell)} \, \\
\delta \!S_{2}^{(\ell)} \!&=&\! \tfrac{1}{2} p \, \Gamma (\epsilon) \left[ \left( B_y + i B_x \right) \Lambda_1 f_{1}^{(\ell)}
+ \left( B_y - i B_x \right) \Lambda_2 f_{3}^{(\ell)} \right]\! , \\
\delta \!S_{-2}^{(\ell)} \!&=&\! \tfrac{1}{2} p \, \Gamma (\epsilon) \left[ \left( B_y - i B_x \right) \Lambda_1 f_{-1}^{(\ell)}
+ \left( B_y + i B_x \right) \Lambda_2 f_{-3}^{(\ell)} \right]\! .
\end{eqnarray*}
Here, $\Gamma (\epsilon)$ is the electronic density of states per spin and per valley, per unit area
$\Gamma = p/(2 \pi \hbar^2 v_g)$,
parameters $\Lambda_1$, $\Lambda_2$ are given by
\begin{eqnarray}
\Lambda_1 \!\! &=& \!\! \Omega (\nu_0 - \nu_2) + \tfrac{1}{2} \Omega_c (\nu_0 - 2\nu_2 + \nu_4) \nonumber \\
&& \qquad + \tfrac{1}{2} \Omega_s (\nu_0 - 2\nu_1 + 2\nu_3 - \nu_4) \, , \label{lam1} \\
\Lambda_2 \!\! &=& \!\! \Omega (\nu_0 + \nu_1 - \nu_2 - \nu_3)
+ \tfrac{1}{2} \Omega_c (\nu_0 - 2\nu_2 - \nu_3 + \nu_4 + \nu_5) \nonumber \\
&& \qquad - \tfrac{1}{2} \Omega_s (\nu_0 - \nu_3 - \nu_4 + \nu_5) \, . \label{lam2}
\end{eqnarray}

The current density is given by~\cite{hei10}
\begin{eqnarray*}
\mathbf{J} = - \frac{g e}{L^2} \sum_{\mathbf{p}} \mathbf{v}_g f (\mathbf{p},t) ,
\end{eqnarray*}
where $g$ is a degeneracy factor ($g=4$ for spin and valley in graphene).
Owing to the angular factors in $\mathbf{v}_g$, only the first order angular harmonics
($m = \pm 1$) in the harmonic expansion~(\ref{expand}) survive after integrating over all angles $\phi$.
We express the current as a series in temporal harmonics as
\begin{eqnarray}
\mathbf{J} &=& \mathbf{J}^{(0)} + \mathbf{J}^{(1)} + \mathbf{J}^{(2)} + \ldots ,\label{jn} \\
\mathbf{J}^{(0)} &=& - \frac{g e}{L^2} \sum_{\mathbf{p}} \mathbf{v}_g
\left( f_1^{(0)} e^{i\phi} + f_{-1}^{(0)} e^{-i\phi} \right) , \nonumber \\
\mathbf{J}^{(n)} &=& - \frac{g e}{L^2} \sum_{\mathbf{p}} \mathbf{v}_g
\Big[ \left( f_1^{(n)} e^{i\phi} + f_{-1}^{(n)} e^{-i\phi} \right)e^{- i n \omega t}   \nonumber \\
&& + \left( f_1^{(-n)} e^{i\phi} + f_{-1}^{(-n)} e^{-i\phi} \right)e^{i n \omega t} \Big] ;
\quad n \geq 1 . \nonumber
\end{eqnarray}
The coupled equations~(\ref{coupled}) are used to express harmonics $f_j^{(\ell)}$
in terms of the equilibrium distribution $f_0^{(0)}$.
Thus, it is possible to calculate each of the harmonic current densities
$\mathbf{J}^{(0)}$, $\mathbf{J}^{(1)}$, $\mathbf{J}^{(2)}$, which we describe below,
beginning with the linear response $\mathbf{J}^{(1)}$.

\subsection{Linear response $\mathbf{J}^{(1)}$}

The leading contribution to the linear current density $\mathbf{J}^{(1)}$ arises
from the linear-in-electric field terms in~(\ref{coupled})
({\em i.e. $\delta \!S_{j}^{(\ell)}$} is irrelevant) with
\begin{eqnarray*}
f_{\pm 1}^{(1)} = \frac{e\tau(E_x \mp iE_y)}{2(1 - i \omega \tau \pm i \omega_c \tau)}
\frac{\partial f_0^{(0)}}{\partial p} ; \qquad
f_{\pm 1}^{(-1)} = \left( f_{\mp 1}^{(1)} \right)^{\ast} .
\end{eqnarray*}
For a degenerate electron gas, $\epsilon_F \gg k_BT$, we find that
$\mathbf{J}^{(1)} = 2 \mathrm{Re} \{ \sigma \mathbf{E} e^{-i\omega t}\}$
where the conductivity tensor $\sigma$ has components as given in Eqs.~(\ref{long},\ref{hall})
with dc Drude conductivity $\sigma_0 = g e^2 v_g p \tau / (4\pi \hbar^2)$
(all parameters are evaluated on the Fermi surface).
For a system with quadratic dispersion $\epsilon = p^2/(2m)$ (such as bilayer graphene),
then $\sigma_0 = g e^2 \epsilon \tau / (2\pi \hbar^2)$ \cite{adamdassarma08},
for linear dispersion $\epsilon = vp$ (monolayer graphene),
then $\sigma_0 = g e^2 \epsilon \tau / (4\pi \hbar^2)$ \cite{and06,nom07,hwa07}.

\subsection{Ratchet dc current $\mathbf{J}^{(0)}$}

The dc current density, Eq.~(\ref{jn}), may be written as
\begin{eqnarray}
\mathbf{J}^{(0)} &=& \Big\{ \left[ \left( \mathbf{B}_{\parallel} \times \mathbf{\hat{n}}_z \right) \cdot \mathbf{E} \right] \mathbf{E}^{\ast}
+ \left[ \left( \mathbf{B}_{\parallel} \times \mathbf{\hat{n}}_z \right) \cdot \mathbf{E}^{\ast} \right] \mathbf{E} \nonumber \\
&& \qquad \qquad
- \left( \mathbf{B}_{\parallel} \times \mathbf{\hat{n}}_z \right) \left| \mathbf{E} \right|^2 \Big\} \mathrm{Re} \left( M_1 \right) \nonumber \nonumber \\
&& + \left( \mathbf{B}_{\parallel} \times \mathbf{\hat{n}}_z \right) \left| \mathbf{E} \right|^2 \mathrm{Re} \left( M_2 \right) \nonumber \\
&& + i \mathbf{B}_{\parallel} \left[ \left( \mathbf{E} \times \mathbf{E}^{\ast} \right) \cdot \mathbf{\hat{n}}_z \right] \mathrm{Re} \left( M_3 \right) \nonumber \\
&& +  \Big\{ \mathbf{E} \left( \mathbf{B}_{\parallel} \cdot \mathbf{E}^{\ast} \right) +
\mathbf{E}^{\ast} \left( \mathbf{B}_{\parallel} \cdot \mathbf{E} \right)
- \mathbf{B}_{\parallel} \left| \mathbf{E} \right|^2 \Big\} \mathrm{Im} \left( M_1 \right) \nonumber \\
&& - \mathbf{B}_{\parallel} \left| \mathbf{E} \right|^2 \mathrm{Im} \left( M_2 \right) \nonumber \\
&& + i \mathbf{B}_{\parallel} \times \left( \mathbf{E} \times \mathbf{E}^{\ast} \right) \mathrm{Im} \left( M_3 \right) \label{j0full}
\end{eqnarray}
In terms of components, this may be expressed~\cite{bud16} as
\begin{eqnarray}
J_x^{(0)} \!\!\!&=&\!\!\! B_x ( - \left| E \right|^2 \mathrm{Im} M_2
+ \Theta_1 \mathrm{Im} M_1 - \Theta_2 \mathrm{Re} M_1  + \Theta_3 \mathrm{Re} M_3 ) \nonumber \\
&& \!\!\! \!\!\! \!\!\! + B_y ( \left| E \right|^2 \mathrm{Re} M_2
+ \Theta_1 \mathrm{Re} M_1 + \Theta_2 \mathrm{Im} M_1  + \Theta_3 \mathrm{Im} M_3 ) , \nonumber \\
J_y^{(0)} \!\!\!&=&\!\!\! B_x ( - \left| E \right|^2 \mathrm{Re} M_2
+ \Theta_1 \mathrm{Re} M_1 + \Theta_2 \mathrm{Im} M_1  - \Theta_3 \mathrm{Im} M_3 ) \nonumber \\
&& \!\!\! \!\!\! \!\!\! + B_y ( - \left| E \right|^2 \mathrm{Im} M_2
- \Theta_1 \mathrm{Im} M_1 + \Theta_2 \mathrm{Re} M_1  + \Theta_3 \mathrm{Re} M_3 ) , \nonumber
\end{eqnarray}
where $\Theta_1 = ( \left| E_x \right|^2-\left| E_y \right|^2)$,
$\Theta_2 = (E_xE_y^*+E_yE_x^*)$
and $\Theta_3 = i (E_xE_y^*-E_yE_x^*)$~\cite{notebud}.

For a degenerate electron gas, $\epsilon_F \gg k_BT$,
we find the three coefficients are given by
\begin{eqnarray}
M_1 &=& - \frac{g e^3}{32 \pi^2 \hbar^4} \left( \frac{1}{\Upsilon^{1,1}} + \frac{1}{\Upsilon^{-1,1}}\right) \nonumber \\
&& \qquad \times
\left[ \frac{4 \Lambda_1 p^2}{\Upsilon^{0,1}\Upsilon^{0,2}}  + v_g p^3  \left( \frac{\Lambda_1}{\Upsilon^{0,1}\Upsilon^{0,2}} \right)^{\prime}\right]  , \nonumber \\
M_2 &=& \frac{g e^3 \Lambda_1 p^2}{32 \pi^2 \hbar^4}
\left( \frac{1}{\Upsilon^{1,2}\Upsilon^{1,1}} + \frac{1}{\Upsilon^{-1,2}\Upsilon^{-1,1}}\right)
\nonumber \\
&& \qquad \times
\left[ \frac{1}{\Upsilon^{0,1}} - \frac{p v_g^{\prime}}{\Upsilon^{0,1}} - v_g p \left( \frac{1}{\Upsilon^{0,1}} \right)^{\prime} \right]  , \nonumber \\
M_3 &=& \frac{i g e^3 \Lambda_1 p^2}{32 \pi^2 \hbar^4}
\left( \frac{1}{\Upsilon^{1,2}\Upsilon^{1,1}} - \frac{1}{\Upsilon^{-1,2}\Upsilon^{-1,1}}\right)
\nonumber \\
&& \qquad \times
\left[ \frac{1}{\Upsilon^{0,1}} - \frac{p v_g^{\prime}}{\Upsilon^{0,1}} - v_g p \left( \frac{1}{\Upsilon^{0,1}} \right)^{\prime} \right]  ,
\label{mcoeffs}
\end{eqnarray}
where $(\ldots)^{\prime} \equiv \partial (\ldots) / \partial \epsilon$ and
all parameters are evaluated on the Fermi surface.
The terms $\mathrm{Re} (M_1)$, $\mathrm{Re} (M_2)$, $\mathrm{Re} (M_3)$ are all
even functions of $\mathbf{\hat{n}}_z \cdot \mathbf{B}_{\perp}$,
whereas $\mathrm{Im} (M_1)$, $\mathrm{Im} (M_2)$, $\mathrm{Im} (M_3)$
are odd functions, thus they are zero for $\mathbf{B}_{\perp}=0$.
These equations generalize those in Refs.~\cite{tara11,drex13,bud16,khe16}
and describe the intraband contribution to the ratchet effect in a two-dimensional material with isotropic dispersion. Parameters such as the scattering times $\tau$, $\tau_2$, Eq.~(\ref{tau}), and $\Lambda_1$, Eq.~(\ref{lam1}), are specific to the given material, we will describe them for bilayer
graphene in Section~\ref{bg}.

The coefficients $M_1$, $M_2$, $M_3$ describe the response
to different polarizations of light: $M_2$ characterizes the effect of unpolarized light,
$M_1$ describes additional terms that appear if the light is linearly polarized,
$M_3$ includes additional terms that occur for circular polarization.
In particular, for incoming linearly-polarized light,
$E_x (t) = E_0 \cos \theta \cos \omega t$ and $E_y (t) = E_0 \sin \theta \cos \omega t$
where $\theta$ is the polarization angle, then
\begin{eqnarray}
E_x^{{\ast}} = E_x = \frac{E_0}{2} \cos \theta \, ; \qquad
E_y^{{\ast}} = E_y = \frac{E_0}{2} \sin \theta \, . \label{lin}
\end{eqnarray}
In this case, the current density may be expressed as
\begin{eqnarray}
\mathbf{J}^{(0)} &=& \frac{E_0^2}{4} B_{\parallel} |M_1| \Big\{
\boldsymbol{\hat{\textbf{\i}}} \cos \left(2\theta - \varphi - \chi_1 + \pi/2\right) \nonumber \\
&& \qquad \qquad \qquad + \boldsymbol{\hat{\textbf{\j}}} \sin \left(2\theta - \varphi - \chi_1 + \pi/2\right) \Big\} , \nonumber \\
&& + \frac{E_0^2}{4} \Big\{ \left( \mathbf{B}_{\parallel} \times \mathbf{\hat{n}}_z \right) \mathrm{Re} M_2
- \mathbf{B}_{\parallel} \mathrm{Im} M_2 \Big\} , \label{j0lin}
\end{eqnarray}
where $\chi_1 = \mathrm{arg} (M_1)$ and
$\varphi$ is the polar angle of the in-plane magnetic field
$\mathbf{B}_{\parallel} = \left( B_x , B_y , 0 \right) = B_{\parallel} \left( \cos \varphi , \sin \varphi , 0 \right)$ and $B_{\parallel} = |\mathbf{B}_{\parallel}|$.
The $M_1$ term produces current in a direction determined by the
polarization angle $\theta$, the magnetic field direction $\varphi$ and the
phase $\chi_1$ of the $M_1$ coefficient, whereas
$M_2$ describes current in a direction solely determined by the parallel field.
For unpolarized light, the $M_1$-related current is zero, but the $M_2$ current
survives.

For circularly-polarized light $E_x (t) = E_0 \cos \omega t$ and $E_y (t) = \mu E_0 \sin \omega t$,
where $\mu = \pm 1$ indicates left- or right-handed polarization,
then
\begin{eqnarray}
E_x^{{\ast}} = E_x = \frac{E_0}{2} \, ; \qquad E_y^{{\ast}} = - E_y = - i \mu \frac{E_0}{2} . \label{cir}
\end{eqnarray}
Then, the dc current density is given by
\begin{eqnarray}
\mathbf{J}^{(0)} &=& \frac{E_0^2}{2} B_{\parallel} |M_2| \Big\{
\boldsymbol{\hat{\textbf{\i}}} \cos \left(\varphi - \chi_2 - \pi/2\right) \nonumber \\
&& \qquad \qquad \qquad + \boldsymbol{\hat{\textbf{\j}}} \sin \left(\varphi - \chi_2 - \pi/2\right) \Big\} , \nonumber \\
&& + \mu \frac{E_0^2}{2} B_{\parallel} |M_3| \Big\{
\boldsymbol{\hat{\textbf{\i}}} \cos \left(\varphi - \chi_3\right) \nonumber \\
&& \qquad \qquad \qquad + \boldsymbol{\hat{\textbf{\j}}} \sin \left(\varphi - \chi_3\right) \Big\} , \label{j0cir}
\end{eqnarray}
where $\chi_2 = \mathrm{arg} (M_2)$, $\chi_3 = \mathrm{arg} (M_3)$,
indicating that the direction of the current is determined by the
magnetic field direction and the phase of the $M_2$, $M_3$ coefficients.

\subsection{Second-harmonic generation $\mathbf{J}^{(2)}$}\label{s:shg}

The second harmonic of the current density, Eq.~(\ref{jn}), may be expressed as
\begin{eqnarray}
\mathbf{J}^{(2)} &=& 2 \mathrm{Re} \Big\{ \left[ \left( \mathbf{B}_{\parallel} \times \mathbf{\hat{n}}_z \right) \mathbf{E}^2 N_3
+ \mathbf{B}_{\parallel} \mathbf{E}^2 N_4 \right] e^{-2i \omega t} \Big\} \nonumber \\
&& \!\!\!\!\!\! \!\!\!\!\!\! \!\!\!\!\!\! + 2 \mathrm{Re} \Big\{ \left[ 2 \left[ \left( \mathbf{B}_{\parallel} \times \mathbf{\hat{n}}_z \right) \cdot \mathbf{E} \right] \mathbf{E}
- \left( \mathbf{B}_{\parallel} \times \mathbf{\hat{n}}_z \right) \mathbf{E}^2 \right] N_1 e^{-2i \omega t} \Big\}  \nonumber \\
&& - 2 \mathrm{Re} \Big\{ \left[ 2 \left( \mathbf{B}_{\parallel} \cdot \mathbf{E} \right) \mathbf{E}
- \mathbf{B}_{\parallel} \mathbf{E}^2 \right] N_2 e^{-2i \omega t} \Big\} , \label{j2full}
\end{eqnarray}
where $\mathbf{E}^2 \equiv \mathbf{E} \cdot \mathbf{E} = E_x^2 + E_y^2$.
In terms of components, this may be written as
\begin{eqnarray*}
J_x^{(2)} &=& 2 \mathrm{Re} \big\{ \big[ N_1 \! \left( B_y \Theta_4 - B_x \Theta_5 \right)
- N_2 \! \left( B_x \Theta_4 + B_y \Theta_5 \right) \nonumber \\
&& \qquad \qquad + \, \Theta_6 \! \left( B_y N_3 + B_x N_4 \right)\big] e^{-2i\omega t} \big\} , \\
J_y^{(2)} &=& 2 \mathrm{Re} \big\{ \big[ N_1 \! \left( B_x \Theta_4 + B_y \Theta_5 \right)
+ N_2 \! \left( B_y \Theta_4 - B_x \Theta_5 \right) \nonumber \\
&& \qquad \qquad + \, \Theta_6 \! \left( - B_x N_3 + B_y N_4 \right)\big] e^{-2i\omega t} \big\} ,
\end{eqnarray*}
where $\Theta_4 = (E_x^2 - E_y^2)$,
$\Theta_5 = 2E_xE_y$
and $\Theta_6 = (E_x^2 + E_y^2)$. For a degenerate electron gas, $\epsilon_F \gg k_BT$,
we find the coefficients $N_i$ are given by
\begin{eqnarray*}
N_1 &=& - \frac{g e^3 v_g p}{32 \pi \hbar^2} \\
&& \times
\left[ \frac{1}{\Upsilon^{1,1}} \left( \frac{v_g p \Gamma \Lambda_1}{\Upsilon^{2,2}\Upsilon^{2,1}} \right)^{\prime}
+ \frac{1}{\Upsilon^{1,-1}} \left( \frac{v_g p \Gamma \Lambda_1}{\Upsilon^{2,-2}\Upsilon^{2,-1}} \right)^{\prime}
\right] , \\
N_2 &=& - \frac{i g e^3 v_g p}{32 \pi \hbar^2} \\
&& \times
\left[ \frac{1}{\Upsilon^{1,1}} \left( \frac{v_g p \Gamma \Lambda_1}{\Upsilon^{2,2}\Upsilon^{2,1}} \right)^{\prime}
- \frac{1}{\Upsilon^{1,-1}} \left( \frac{v_g p \Gamma \Lambda_1}{\Upsilon^{2,-2}\Upsilon^{2,-1}} \right)^{\prime}
\right] , \\
N_3 &=& \frac{g e^3 v_g p \Gamma \Lambda_1}{32 \pi \hbar^2}
\Bigg[ \frac{1}{\Upsilon^{1,1}\Upsilon^{1,2}\Upsilon^{2,1}} + \frac{1}{\Upsilon^{1,-1}\Upsilon^{1,-2}\Upsilon^{2,-1}} \\
&& - \frac{p}{\Upsilon^{1,1}\Upsilon^{1,2}} \left( \frac{v_g}{\Upsilon^{2,1}} \right)^{\prime}
- \frac{p}{\Upsilon^{1,-1}\Upsilon^{1,-2}} \left( \frac{v_g}{\Upsilon^{2,-1}} \right)^{\prime} \Bigg] , \\
N_4 &=& \frac{i g e^3 v_g p \Gamma \Lambda_1}{32 \pi \hbar^2}
\Bigg[ \frac{1}{\Upsilon^{1,1}\Upsilon^{1,2}\Upsilon^{2,1}} - \frac{1}{\Upsilon^{1,-1}\Upsilon^{1,-2}\Upsilon^{2,-1}} \\
&& - \frac{p}{\Upsilon^{1,1}\Upsilon^{1,2}} \left( \frac{v_g}{\Upsilon^{2,1}} \right)^{\prime}
+ \frac{p}{\Upsilon^{1,-1}\Upsilon^{1,-2}} \left( \frac{v_g}{\Upsilon^{2,-1}} \right)^{\prime} \Bigg] .
\end{eqnarray*}
Note that coefficients $N_1$, $N_3$ are even functions of $\mathbf{\hat{n}}_z \cdot \mathbf{B}_{\perp}$,
whereas $N_2$, $N_4$ are odd, thus $N_2 = N_4 = 0$ for $\mathbf{B}_{\perp} = 0$.

For incoming linearly-polarized light Eq.~(\ref{lin}), the current density is
\begin{eqnarray}
\mathbf{J}^{(2)} &=& \frac{E_0^2}{2} B_{\parallel} |N_1| \cos \left( 2 \omega t - \psi_1 \right)
\Big\{
\boldsymbol{\hat{\textbf{\i}}} \cos \left(2\theta - \varphi + \pi/2\right) \nonumber \\
&& \qquad \qquad \qquad + \boldsymbol{\hat{\textbf{\j}}} \sin \left(2\theta - \varphi + \pi/2\right) \Big\} , \nonumber \\
&& + \frac{E_0^2}{2} B_{\parallel} |N_2| \cos \left( 2 \omega t - \psi_2 \right)
\Big\{
\boldsymbol{\hat{\textbf{\i}}} \cos \left(2\theta - \varphi + \pi \right) \nonumber \\
&& \qquad \qquad \qquad + \boldsymbol{\hat{\textbf{\j}}} \sin \left(2\theta - \varphi + \pi\right) \Big\} , \nonumber \\
&& + \frac{E_0^2}{2} |N_3| \left( \mathbf{B}_{\parallel} \times \mathbf{\hat{n}}_z \right)
\cos \left( 2 \omega t - \psi_3 \right) , \nonumber \\
&& + \frac{E_0^2}{2} |N_4| \mathbf{B}_{\parallel}
\cos \left( 2 \omega t - \psi_4 \right) , \label{j2lin}
\end{eqnarray}
where the phases $\psi_1 = \mathrm{arg} (N_1)$, $\psi_2 = \mathrm{arg} (N_2)$, $\psi_3 = \mathrm{arg} (N_3)$,
$\psi_4 = \mathrm{arg} (N_4)$ describe a time lag between the incoming light and the produced current.
For the $N_1$ and $N_2$ terms, the in-plane magnetic field rotates the polarization direction
as in the Faraday effect~\cite{ped96} whereas, for the $N_3$ and $N_4$ terms, the outgoing linear polarization
direction is solely determined by the parallel field (it is independent of the incoming
polarization direction $\theta$).
Note that, for unpolarized light, the $N_1$ and $N_2$-related currents are zero, but the $N_3$ and $N_4$ currents
survive.

For incoming circularly-polarized light Eq.~(\ref{cir}), the
$N_3$ and $N_4$-related currents are zero, and the current density is
\begin{eqnarray}
\mathbf{J}^{(2)} &=& E_0^2 B_{\parallel} |N_1|
\Big\{
\boldsymbol{\hat{\textbf{\i}}} \cos \left(2 \omega t - \psi_1 - \mu [\varphi - \pi/2]\right) \nonumber \\
&& \qquad \qquad + \mu \boldsymbol{\hat{\textbf{\j}}} \sin \left(2 \omega t - \psi_1 - \mu [\varphi - \pi/2]\right) \Big\} , \nonumber \\
&& + E_0^2 B_{\parallel} |N_2|
\Big\{
\boldsymbol{\hat{\textbf{\i}}} \cos \left(2 \omega t - \psi_2 - \mu \varphi - \pi\right) \nonumber \\
&& \qquad \qquad + \mu \boldsymbol{\hat{\textbf{\j}}} \sin \left(2 \omega t - \psi_2 - \mu \varphi - \pi\right) \Big\} . \label{j2cir}
\end{eqnarray}
Thus, the generated current is also circularly polarized, the direction of the
magnetic field contributes to the phase lag.

\section{Bilayer graphene}\label{bg}

\subsection{Four-component Hamiltonian}

\begin{figure}[ht]
   \centering
   \includegraphics[scale=0.35]{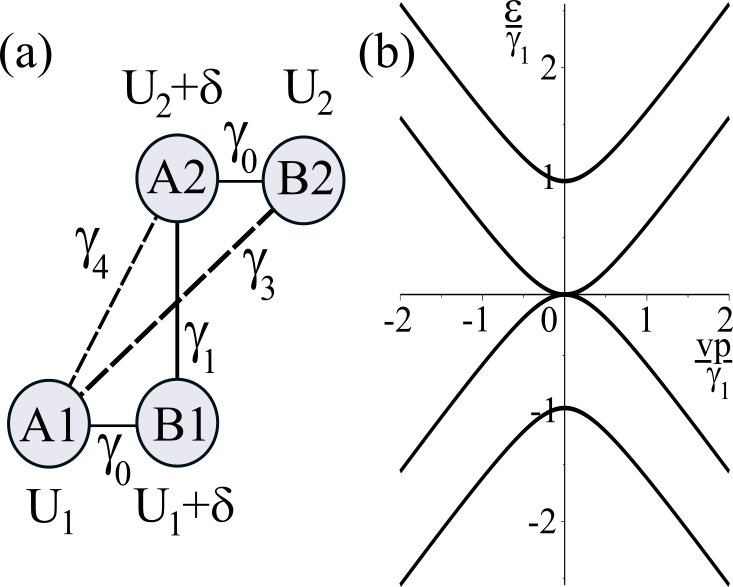}
   \caption{(a) Schematic of the unit cell of bilayer graphene with four atoms in the unit
   cell, A1, B1 on the lower layer, A2, B2 on the upper layer, and relevant
   tight-binding parameters. (b) Schematic of the four energy bands near each K point including two conduction bands
    and two valence bands.
   }
    \label{fig:lat}
\end{figure}

In order to apply the general equations for the second-order conductivities
derived in Section~\ref{s:be} to a particular system, it is necessary to model
scattering in the presence of an in-plane magnetic field in that system in order to
derive the form of parameters $\Lambda_1$ [Eq.~(\ref{lam1})]
and $\Omega$, $\Omega_c$, $\Omega_s$ [Eq.~(\ref{wgen})].
For bilayer graphene, this has been done previously~\cite{khe16} in order to
model the magnetic ratchet, here we also include cyclotron resonance and
second harmonic generation. We will briefly describe
electronic scattering in bilayer graphene in the presence of an in-plane magnetic field
but we refer the reader to~\cite{khe16} for further details.

Bilayer graphene has four atomic sites in the unit cell, Fig.~\ref{fig:lat}(a), we label them as
A1, B1 on the lower layer, and A2, B2 on the upper layer.
Sites B1 and A2 lie directly below and above each other, and, as a result, their orbitals
are relatively-strongly coupled and these sites are referred to as `dimer' sites.
We employ a Cartesian coordinate system with the graphene lying in the $x$-$y$ plane,
$z$ in the perpendicular direction, the lower layer
of the bilayer is at $z = -d/2$, the upper layer at $z= d/2$, where $d$ is the interlayer spacing.
We use the tight-binding model \cite{mccfal06,gui06,nil08,mcckos13}
with one $p_z$ orbital per site,
and we take the in-plane magnetic field into account
$\mathbf{B}_{\parallel} = (B_x,B_y,0)$
with a vector potential $\mathbf{A} = z ( B_y , -B_x , 0)$ that preserves
translational symmetry in the $x$-$y$ graphene plane.

The vector potential enters the model through a line integral
appearing in the matrix elements, for example,
the matrix element for in-plane hopping between an $A$ atom at $\mathbf{R}_{A}$
and three nearest-neighbor $B$ atoms at $\mathbf{R}_{Bj}$, $j=1,2,3$, is given by
\begin{eqnarray*}
H_{AB} = - \gamma_0 \sum_{j = 1}^3 \exp \! \left(i \mathbf{k} \cdot \left( \mathbf{R}_{Bj} - \mathbf{R}_{A}\right) - \frac{i e }{\hbar}\int_{\mathbf{R}_{Bj}}^{\mathbf{R}_{A}}
\mathbf{A} . \mathbf{d \ell} \right)  ,
\end{eqnarray*}
where $\gamma_0$ is a tight-binding parameter and $\mathbf{k}$ is the wave vector.
Two non-equivalent valleys are located at the Brillouin zone corners (K points), wave vector $\mathbf{K}_{\xi} = \xi (4\pi/3a, 0)$,
$\xi = \pm 1$, and, in the vicinity of these points, the in-plane momentum is
$\mathbf{p}=  ( p_x , p_y , 0) = \hbar \mathbf{k} - \hbar \mathbf{K}_{\xi}$.
Keeping linear in $\mathbf{p}$ and linear in $\mathbf{B}_{\parallel}$ contributions,
the Hamiltonian~\cite{khe16} in a basis of A1, B1, A2, B2 sites is
\begin{eqnarray}
H = \left(
      \begin{array}{cccc}
        U_1 & v\pi_1^{\dagger} & - v_4\pi^{\dagger} & v_3\pi \\
        v\pi_1 & U_1+\delta & \gamma_1 & - v_4\pi^{\dagger} \\
        - v_4\pi & \gamma_1 & U_2+\delta & v\pi_2^{\dagger} \\
        v_3\pi^{\dagger} & - v_4\pi & v\pi_2 & U_2 \\
      \end{array}
    \right) \, , \label{h4}
\end{eqnarray}
where $v = \sqrt{3} a \gamma_0 / (2\hbar)$ represents in-plane
nearest-neighbour A1-B1, A2-B2 hopping, $a$ is the lattice constant,
$\gamma_1$ describes vertical interlayer coupling,
$v_3 = \sqrt{3} a \gamma_3 / (2\hbar)$ represents skew interlayer
A1-B2 hopping, and
$v_4 = \sqrt{3} a \gamma_4 / (2\hbar)$ represents skew interlayer
A1-A2, B1-B2 hopping, Fig.~\ref{fig:lat}(a).
The on-site energies of the two layers are characterized by $U_1$, $U_2$,
while $\delta$ describes an energy difference between B1, A2 (dimer sites)
and A1, B2 (non-dimers) \cite{dressel02,nil08,zhang08,li09,mcckos13}.
Complex momentum operators are labelled $\pi_1$ for the lower layer,
$\pi_2$ for the upper layer and $\pi$ for interlayer hopping:
\begin{eqnarray*}
\pi &=& \xi p_x + i p_y \, , \\
\pi_1 &=& \xi ( p_x - b_y ) + i ( p_y + b_x ) \, , \\
\pi_2 &=& \xi ( p_x + b_y ) + i ( p_y - b_x ) \, ,
\end{eqnarray*}
where the magnetic field, written in dimensions of momentum, is $b_x = e d B_x /2$, $b_y = e d B_y / 2$.

\subsection{Two-component reduced low-energy Hamiltonian}

There are four $p_z$ orbitals in the unit cell and Hamiltonian~(\ref{h4})
describes four energy bands near each K point, two conduction bands, two valence bands, Fig.~\ref{fig:lat}(b).
Of these, one of the conduction bands touches one valence band at the K point with
an approximately quadratic dispersion $\epsilon = v^2 p^2/\gamma_1$
near zero energy ~\cite{mccfal06,gui06,nil08,mcckos13}.
The other two bands are split away from the touching point by $\pm \gamma_1$
because the orbitals corresponding to the B1, A2 (dimer) sites are strongly coupled
by $\gamma_1$. Thus, it is possible to represent the electronic behavior at low
energy (less than $|\gamma_1|$) where there are only two bands by eliminating
the components in Hamiltonian~(\ref{h4}) related to B1, A2 (dimer) sites,
resulting in a two-component Hamiltonian for the A1, B2 (non-dimer) sites.
This process has been described before~\cite{mccfal06,mcckos13},
and including the in-plane magnetic field~\cite{khe16}, the two-component
Hamiltonian in an A1, B2 basis is
\begin{eqnarray}
H &=& - \frac{v^2}{\gamma_1} \left(
                             \begin{array}{cc}
                               0 & \left(\pi^{\dagger}\right)^2 \\
                               \pi^2 & 0 \\
                             \end{array}
                           \right) \nonumber \\
&& \, + \frac{\Delta}{2} \!\left[ 1 - \frac{2v^2p^2}{\gamma_1^2}\right]\! \left(
       \begin{array}{cc}
         1 & 0 \\
         0 & -1 \\
       \end{array}
     \right)  \, \nonumber \\
&& \, - \frac{2 v^2}{\gamma_1} \!\left[ \frac{v_4}{v} + \frac{\delta}{\gamma_1}\right]\! ( \mathbf{p} \times \mathbf{b} )_z \left(
                           \begin{array}{cc}
                             1 & 0 \\
                             0 & -1 \\
                           \end{array}
                         \right) \, \nonumber \\
&& \, - \frac{v v_4 \Delta}{\gamma_1^2} \left(
                           \begin{array}{cc}
                           0 & i\pi^{\dagger} \beta^{\dagger} \\
                           -i \pi \beta & 0 \\
                           \end{array}
                           \right)  \label{ham3} ,
\end{eqnarray}
where $\beta = b_x + i \xi b_y$, $\beta^{\dagger} = b_x - i \xi b_y$, $p = |\mathbf{p}|$,
and interlayer asymmetry is $\Delta = U_1 - U_2$.
To derive Eq.~(\ref{ham3}), we neglected a number of contributions, further details may be found in
Ref.~\cite{khe16}. We excluded
terms that are quadratic or higher in the magnetic field, cubic or higher in $vp/\gamma_1$ and
cubic or higher in other small parameters $v_4/v$, $\delta / \gamma_1$,
$U_1 / \gamma_1$ and $U_2 / \gamma_1$.
Note that tight-binding models that neglect parameters $v_4$ and $\delta$
omit the linear-in-$\mathbf{B}_{\parallel}$ terms~\cite{persh10,hyu12,gon12,roy13,he14,don16,don16c}.
Additionally, we neglect terms that are proportional to the unit matrix in (A1, B2) space because
they don't influence electronic scattering despite having a small effect on the dispersion relation.

In Eq.~(\ref{ham3}), we neglect parameter $\gamma_3$. It doesn't produce magnetic field dependent terms
in the two-component Hamiltonian so that, for the magnetic ratchet effect, it only contributes to small cross terms in the
scattering probability that are higher order in small parameters than the results quoted here.
It is reasonable to wonder whether $\gamma_3$ and interlayer asymmetry $\Delta$ could
produce second-order-in-$\mathbf{E}$ currents for $\mathbf{B}_{\parallel} = 0$. In fact, $\Delta$
on its own gives only isotropic terms in Eq.~(\ref{ham3}), thus
it doesn't create coupling between harmonics, it just
gives a small correction to the momentum scattering times $\tau$, $\tau_2$.
Parameter $\gamma_3$ does produce anisotropic terms in the two-component Hamiltonian
which can lead to coupling between different harmonics.
However, even in the presence of finite $\Delta$, the
sign of the resulting second-order-in-$\mathbf{E}$ current
is valley dependent (like the orientation of $\gamma_3$-induced trigonal warping
of the band structure \cite{mccfal06}) and,
after summing over both valleys, the resulting current is zero.
This valley current is described in detail in Section~\ref{s:valley}.

The first term in the Hamiltonian~(\ref{ham3}) represents chiral quasiparticles~\cite{novo06,mccfal06} with the pseudospin direction [the relative amplitude of the electronic wave function in the lattice (A1, B2) space] in the graphene plane and fixed to the direction of the electronic momentum $\mathbf{p} =  ( p_x , p_y , 0)$. This term results in a quadratic dispersion $\epsilon = v^2 p^2/\gamma_1$,
and we assume that the other terms in~(\ref{ham3}) are a small perturbation with respect to this dominant one.
The second term in~(\ref{ham3}) describes a gap in the spectrum~\cite{mccfal06} due to different energies $U_1$, $U_2$ on the two layers as characterized by interlayer asymmetry $\Delta = U_1 - U_2$.
Such interlayer asymmetry could be induced by an external gate voltage and, thus, parameter $\Delta$
is, in principle, tuneable. Also, the
presence of a substrate could break inversion symmetry by creating a different
electrostatic potential in one layer of the bilayer as compared to the other.
Substrate-induced $\Delta$ could be significant in certain circumstances: for example, it has been estimated to be
$30\,$meV for rippled graphene on SiO$_2$ \cite{kim08} and for graphene on hexagonal boron nitride at a small misalignment angle \cite{yan12,moon14,chen16}.

The in-plane magnetic field appears in two different terms.
The first [third term in~(\ref{ham3})] arises due to small, intrinsic lattice parameters $v_4$ and
$\delta$, and it tends to open a gap with a direction dependent on
the $z$ component of the Lorentz force $\mathbf{p} \times \mathbf{B}_{\parallel}$.
The magnetic field also appears in the fourth term in~(\ref{ham3}), this term only appears
when there is non-zero interlayer asymmetry $\Delta$.
Next, we use Hamiltonian~(\ref{ham3}) to calculate the correction to the
scattering rate~(\ref{formulsp}) due to the in-plane field.

\subsection{Electron scattering}

We determine the scattering rate~(\ref{formulsp}) using
Hamiltonian~(\ref{ham3}) to determine eigenstates $|\mathbf{p}\rangle$ and $|\mathbf{p}^{\prime}\rangle$
in the presence of the in-plane field $\mathbf{B}_{\parallel}$,
with scattering $\delta H$ caused by static impurities~(\ref{static}).
The dimensionless matrix $\hat{Y}$ in $\delta H$ accounts for
structure in the A1, B2 lattice degrees of freedom,
that is, a possible asymmetry in the level of disorder on the two layers
of the bilayer.
As representative examples, we consider disorder that is symmetric with equal amounts of scattering
on the two layers, $\hat{Y} = \hat{I}$ where $\hat{I}$ is the unit matrix,
and we consider asymmetric disorder,
$\hat{Y} = (\hat{I} + \zeta\hat{\sigma}_z)/2$, with scattering limited to the lower ($\zeta = 1$)
or upper ($\zeta = -1$) layer.
Thus, the linear-in-$\mathbf{B}_{\parallel}$ part of the scattering rate
may be written~\cite{khe16} in a general form as in Eq.~(\ref{wgen})
where the angle-independent factors $\Omega$, $\Omega_c$, $\Omega_s$ are:
\begin{eqnarray}
\Omega_c^{(\mathrm{s})} = \Omega^{(\mathrm{s})}
&=& \frac{\pi e d n_i \Delta \gamma_1}{2\hbar v^2 p^4} \!\!
\left( \frac{\gamma_4}{\gamma_0} + \frac{\delta}{\gamma_1} \right) \!\!
\left( 1 - \frac{2v^2p^2}{\gamma_1^2} \right) \, , \nonumber \\
\Omega_s^{(\mathrm{s})} &=& \frac{\pi e d n_i \Delta \gamma_4}{2\hbar p^2 \gamma_1 \gamma_0} \, , \label{scatts}
\end{eqnarray}
for symmetric disorder, and for asymmetric disorder:
\begin{eqnarray}
\Omega^{(\mathrm{a})}
&=& \frac{\pi e d n_i }{2\hbar p^2} \!\!
\left( \frac{\gamma_4}{\gamma_0} + \frac{\delta}{\gamma_1} \right) \!\!
\left( s \zeta - \frac{\Delta}{\gamma_1} + \frac{\Delta \gamma_1}{2v^2p^2} \right) \, , \nonumber \\
\Omega_c^{(\mathrm{a})} = \Omega_s^{(\mathrm{a})} &=& 0 \, , \label{scatta}
\end{eqnarray}
where the density of impurities is $n_i = N_i/L^2$,
and $s=+1$ ($s=-1$) for states in the conduction (valence) band.
Clearly, for symmetric disorder, there must be
interlayer asymmetry $\Delta = U_1 - U_2$ to create $\mathbf{B}_{\parallel}$-dependent terms,
but, for asymmetric disorder, this is not necessary.

\subsection{Ratchet dc current $\mathbf{J}^{(0)}$ in bilayer graphene}\label{s:rat}

For bilayer graphene, we assume that the dispersion is quadratic~\cite{mccfal06}
$\epsilon = v^2p^2/\gamma_1 \equiv p^2/(2m)$ with mass $m = \gamma_1 / 2v^2$ and $v_g = 2v^2p/\gamma_1$.
Then, the factor $1 - p v_g^{\prime} - v_g p \tau^{\prime}/\tau$ in coefficients $M_2$ and $M_3$,
Eq.~(\ref{mcoeffs}), simplifies as $- v_g p \tau^{\prime}/\tau$
and these coefficients are zero unless $\tau$ is energy dependent.
Overscreened Coulomb impurities in bilayer graphene act like short-range scatterers \cite{adamdassarma08},
$u \! \left( \mathbf{r} - \mathbf{R}_j \right) = u_0 \delta \! \left( \mathbf{r} - \mathbf{R}_j \right)$
and $\tilde{u} \! \left( \mathbf{p}^{\prime} - \mathbf{p} \right) = u_0$,
and the scattering rates Eq.~(\ref{tau}) are
\begin{eqnarray*}
\mathrm{symmetric\,\,disorder\!:} \quad \tau^{-1} &=& 2 \tau_2^{-1} = \frac{n_i u_0^2 \gamma_1}{4 \hbar^3 v^2} \, , \\
\mathrm{asymmetric\,\,disorder\!:} \quad \tau^{-1} &=& \tau_2^{-1} = \frac{n_i u_0^2 \gamma_1}{8 \hbar^3 v^2} \, . \\
\end{eqnarray*}
Then, if $u_0$ is independent of energy, so is $\tau$ and $M_2 = M_3 = 0$.
The potential is isotropic so the only non-zero harmonic is $\nu_0 = u_0^2$ and
parameter $\Lambda_1$, Eq.~(\ref{lam1}), becomes
$\Lambda_1 = u_0^2 [ \Omega + (\Omega_c + \Omega_s)/2]$.
Then, the non-linear coefficients, $M_1^{(\mathrm{s})}$
for symmetric disorder Eq.~(\ref{scatts}) and
$M_1^{(\mathrm{a})}$ for asymmetric disorder Eq.~(\ref{scatta}), are
\begin{eqnarray}
M_1^{(\mathrm{s})} &=& \frac{e^4 d}{8\pi \hbar^2 m \tau} \frac{\Delta}{\gamma_1}
\left(\frac{5\gamma_4}{\gamma_0} +\frac{6\delta}{\gamma_1} \right) \nonumber \\
&& \qquad \times \frac{1}{\Upsilon^{0,1}\Upsilon^{0,2}}
\left( \frac{1}{\Upsilon^{1,1}} + \frac{1}{\Upsilon^{-1,1}}\right) , \label{ms} \\
M_1^{(\mathrm{a})} &=& - \frac{e^4 d}{2\pi \hbar^2 m \tau} \!\!
\left( \frac{\gamma_4}{\gamma_0} + \frac{\delta}{\gamma_1} \right)\!\!\!\left( s \zeta - \frac{\Delta}{\gamma_1} \right) \nonumber \\
&& \qquad \times \frac{1}{\Upsilon^{0,1}\Upsilon^{0,2}}
\left( \frac{1}{\Upsilon^{1,1}} + \frac{1}{\Upsilon^{-1,1}}\right) , \label{ma}
\end{eqnarray}
which generalize the results for $\omega_c = 0$ from Ref.~\cite{khe16}.
Note that both coefficients change sign under $z \rightarrow - z$ inversion
because of the presence of $\Delta$ (interlayer asymmetry)
and $\zeta$ (asymmetric disorder),
and either mechanism (interlayer asymmetry or asymmetric disorder)
produces non-zero $M_1$.

\subsection{Second-harmonic generation $\mathbf{J}^{(2)}$ in bilayer graphene}

For quadratic dispersion and energy-independent scattering rates, $N_3 = N_4 = 0$
and the second harmonic is given by Eq.~(\ref{j2intro}) with
\begin{eqnarray}
N_1^{(\mathrm{s,a})} &=& \frac{c^{(\mathrm{s,a})} e^4 d}{\pi \hbar^2 m \tau} \frac{\Delta}{\epsilon}
\left( \frac{\gamma_4}{\gamma_0} + \frac{\delta}{\gamma_1} \right) \nonumber \\
&& \!\!\! \!\!\! \!\!\! \times \!
\left( \frac{1}{\Upsilon^{1,1}\Upsilon^{2,2}\Upsilon^{2,1}}
+ \frac{1}{\Upsilon^{1,-1}\Upsilon^{2,-2}\Upsilon^{2,-1}} \right) \! , \label{n1f} \\
N_2^{(\mathrm{s,a})} &=& \frac{i c^{(\mathrm{s,a})} e^4 d}{\pi \hbar^2 m \tau} \frac{\Delta}{\epsilon}
\left( \frac{\gamma_4}{\gamma_0} + \frac{\delta}{\gamma_1} \right) \nonumber \\
&& \!\!\! \!\!\! \!\!\! \times \!
\left( \frac{1}{\Upsilon^{1,1}\Upsilon^{2,2}\Upsilon^{2,1}}
- \frac{1}{\Upsilon^{1,-1}\Upsilon^{2,-2}\Upsilon^{2,-1}} \right) \! , \label{n2f}
\end{eqnarray}
where the numerical factor is $c^{(\mathrm{s})} = 3/16$ for symmetric disorder, Eq.~(\ref{scatts}),
and $c^{(\mathrm{a})} = 1/8$ for asymmetric disorder, Eq.~(\ref{scatta}).
For the case of asymmetric disorder, there is no contribution
that is independent of $\Delta$, {\em i.e.} asymmetric disorder on its own cannot produce
the second harmonic in bilayer graphene, in contrast to the dc current $M_1^{(\mathrm{a})}$.

As discussed in Section~\ref{s:shg}, the $N_1$ and $N_2$-related currents are zero for incoming
unpolarized light, whereas, for incoming linear polarization, Eq.~(\ref{j2lin}),
the in-plane magnetic field rotates the polarization direction
as in the Faraday effect~\cite{ped96}.
For incoming circularly-polarized light, Eq.~(\ref{j2cir}),
the generated current is also circularly polarized and the direction of the
magnetic field contributes to the phase lag.

\subsection{Discussion}

The magnitude of the ratchet current in bilayer graphene may be estimated using values of tight-binding parameters
determined by infrared spectroscopy~\cite{zhang08} (see also Ref.~\cite{kuz09}) such as
$\gamma_0 = 3.0\,$eV, $\gamma_1 = 0.4\,$eV, $\gamma_4 = 0.015\,$eV, $\delta = 0.018\,$eV.
We also use interlayer spacing $d \approx 3.3\,${\AA} and mass $m \approx 0.05m_e$ where $m_e$ is the free electron mass.
For a typical value $\tau = 0.15\,$ps~\cite{gor07}
and with values $|\mathbf{E}| = 10\,$kV\,cm$^{-1}$, $|\mathbf{B}| = 7\,$T, $\omega = 2.1 \times 10^{13}\,$rad\,s$^{-1}$ from recent experiment~\cite{drex13}, we estimate that the ratchet current density for asymmetric disorder Eq.~(\ref{ma}) is of the order of $|\mathbf{J}| \sim |M_1^{(\mathrm{a})}| |\mathbf{B}| |\mathbf{E}|^2 \sim 1\,$mA\,cm$^{-1}$.
Note, however, that these quoted experimental values violate the condition of validity
of the Boltzmann equation, $e |\mathbf{E}| v_g \tau \ll k_B T$, and this should only
be considered as an order of magnitude estimation.
The second harmonic, Eqs.~(\ref{n1f},\ref{n2f}), is generally of the same order of magnitude, but
with an additional small parameter $\Delta / \epsilon_F$, its precise value is not fixed
because $\epsilon_F$ and $\Delta$ are both tunable.

A dc current may also be generated by photogalvanics or photon-drag effects \cite{jiang11}
or, in small samples with sufficiently high mobility, by edge photogalvanics \cite{karch11}.
The ratchet current described here could be distinguished by its dependence on electric
and magnetic field directions Eq.~(\ref{j0full}) or the frequency dependence of the
$M_1$, $M_2$, $M_3$ coefficients Eq.~(\ref{mcoeffs}).

Our results apply to intraband transitions in the semiclassical regime $\epsilon_F \gg \omega$
(note that recent papers~\cite{wu12,brun15} consider second harmonic generation in bilayer
graphene at higher frequencies).
For Fermi energy $\epsilon_F = 100\,$meV, say, the frequency corresponding
to $\hbar \omega = 100\,$meV is $\omega \approx 150 \times 10^{12}\,$rad$\,$s$^{-1}$
or linear frequency $f \approx 24\,$THz.
For $\omega_c = eB_{\perp}/m$ with $m \approx 0.05m_e$ for bilayer graphene, where $m_e$ is the free electron mass,
then a perpendicular field $B_{\perp} = 1\,$T corresponds to
$\omega_c \approx 3.5 \times 10^{12}\,$rad$\,$s$^{-1}$.
Then, the cyclotron resonance condition $\omega = \omega_c$ corresponds to
linear frequency of light $f \approx 0.56\,$THz
(the corresponding energy scale is $\hbar \omega \approx 15\,$meV)
which is well within the semiclassical regime considered here.

\section{Valley currents in bilayer graphene for $\mathbf{B}_{\parallel} = 0$
due to trigonal warping}\label{s:valley}

Here we consider second-order-in-electric-field
currents that occur when $\mathbf{B}_{\parallel} = 0$ due to the presence of
skew interlayer A1-B2 hopping described by parameter $\gamma_3$, Eq.~(\ref{h4}).
These are valley currents: the sign
of the current per valley depends on the valley index ($\xi = \pm 1$) and
the total current, summed over both valleys, is zero unless there is an additional
source of valley polarization \cite{ryc07,xiao07,wu12,wang14,yu14,gol14,brun15}.
The two-component Hamiltonian~(\ref{ham3}) is modified as
\begin{eqnarray}
H &=& - \frac{v^2}{\gamma_1} \left(
                             \begin{array}{cc}
                               0 & \left(\pi^{\dagger}\right)^2 \\
                               \pi^2 & 0 \\
                             \end{array}
                           \right)
+ v_3 \left(
                             \begin{array}{cc}
                               0 & \pi \\
                               \pi^{\dagger} & 0 \\
                             \end{array}
                           \right) \nonumber \\
&& \, + \frac{\Delta}{2} \!\left[ 1 - \frac{2v^2p^2}{\gamma_1^2}\right]\! \left(
       \begin{array}{cc}
         1 & 0 \\
         0 & -1 \\
       \end{array}
     \right)  , \label{ham4}
\end{eqnarray}
where $v_3 = \sqrt{3} a \gamma_3 / (2\hbar)$. The presence of $\gamma_3$
creates trigonal warping of the electron band structure \cite{mccfal06,mccab07,mcckos13}
in which the Fermi circle around each valley assumes a triangular distortion,
with the orientation of the distortion being opposite in the two valleys.

As in Section~\ref{bg}, we assume that the first term in the Hamiltonian~(\ref{ham4})
dominates, and the other terms are small with respect to it,
{\em i.e.} $v^2 p^2/\gamma_1 \gg \{ v_3 p , \Delta \}$.
Interlayer asymmetry $\Delta$ has no angular dependence in~(\ref{ham4})
and it doesn't create coupling between harmonics~Eq.(\ref{coupled}), it just gives a small
correction to the momentum scattering times $\tau$, $\tau_2$ which we neglect.
Parameter $\gamma_3$, however, does produce coupling between
different harmonics. To lowest order in $\gamma_3$ the relevant
correction to the scattering rate for symmetric disorder, $\hat{Y} = \hat{I}$, is
\begin{eqnarray*}
\delta W_{\mathbf{p}^{\prime}\mathbf{p}}^{(\mathrm{s})} &=& \xi \frac{\pi}{\hbar}
\frac{n_i}{L^2} \left| \tilde{u} \! \left( \mathbf{p}^{\prime} - \mathbf{p} \right) \right|^2
\delta \! \left( \epsilon_{\mathbf{p}} - \epsilon_{\mathbf{p}^{\prime}} \right)  \\
&& \times \frac{v_3 \gamma_1}{v^2 p}
\bigg\{\cos [2(\phi^{\prime} - \phi)] \left[ \cos (3\phi) + \cos (3\phi^{\prime}) \right] \\
&& \quad - 2 \cos [\tfrac{1}{2}(\phi^{\prime} - \phi)]\cos [\tfrac{3}{2}(\phi^{\prime} + \phi)]
\bigg\}  , \nonumber
\end{eqnarray*}
and for asymmetric disorder, $\hat{Y} = (\hat{I} + \zeta\hat{\sigma}_z)/2$, it is
\begin{eqnarray*}
\delta W_{\mathbf{p}^{\prime}\mathbf{p}}^{(\mathrm{a})} &=&
s \zeta \frac{\pi n_i\gamma_1 \Delta}{4\hbar v^2 L^2}\left( 1 - \tfrac{2v^2p^2}{\gamma_1^2}\right)
\left| \tilde{u} \! \left( \mathbf{p}^{\prime} - \mathbf{p} \right) \right|^2
\delta \! \left( \epsilon_{\mathbf{p}} - \epsilon_{\mathbf{p}^{\prime}} \right)  \\
&& \!\!\!\!\!\! \!\!\!\!\!\! \times
\left[ \frac{1}{p^2} + \frac{1}{(p^{\prime})^2} +
\frac{\xi v_3 \gamma_1}{v^2p^3}\cos (3\phi) +
\frac{\xi v_3 \gamma_1}{v^2(p^{\prime})^3}\cos (3\phi^{\prime}) \right] .
\end{eqnarray*}
Following the harmonic expansion, these terms lead to coupling between
different harmonics Eq.(\ref{coupled}) with
\begin{eqnarray}
\delta S_j^{(\ell)} = C_{j-3} f_{j-3}^{(\ell)} + C_{-j-3} f_{j+3}^{(\ell)} \, , \label{wcouple}
\end{eqnarray}
where, for symmetric or asymmetric disorder,
\begin{eqnarray}
C_{j}^{(\mathrm{s})} &=& \xi \frac{\pi n_i \Gamma v_3 \gamma_1}{4\hbar v^2 p}  \\
&& \!\!\! \times \!\left( \nu_{j+5} - \nu_{5} - \nu_{j+2} + \nu_{j-2} - \nu_{j+1} + \nu_{1} \right)  , \nonumber \\
C_{j}^{(\mathrm{a})} &=& s \xi \zeta \frac{\pi n_i \Gamma v_3 \Delta \gamma_1^2}{4\hbar v^4 p^3}
\left( 1 - \tfrac{2v^2p^2}{\gamma_1^2}\right) \left( \nu_{j} - \nu_{0} \right)  .
\end{eqnarray}

The dc current per valley $\mathbf{J}^{(0)}$ and the
second harmonic current per valley $\mathbf{J}^{(2)}$ may be written as
\begin{eqnarray}
\mathbf{J}^{(0)} &=& \left( \boldsymbol{\hat{\textbf{\i}}} \Theta_1 - \boldsymbol{\hat{\textbf{\j}}} \Theta_2 \right) \! \mathrm{Re} \:\! Q
- \left( \boldsymbol{\hat{\textbf{\i}}} \Theta_2 + \boldsymbol{\hat{\textbf{\j}}} \Theta_1 \right) \! \mathrm{Im} \:\!  Q , \label{dcw} \\
\mathbf{J}^{(2)} &=& 2 \mathrm{Re} \! \left\{ \left[
S_1 \!\left( \boldsymbol{\hat{\textbf{\i}}} \Theta_4 - \boldsymbol{\hat{\textbf{\j}}} \Theta_5 \right)
+ S_2 \! \left( \boldsymbol{\hat{\textbf{\i}}} \Theta_5 + \boldsymbol{\hat{\textbf{\j}}} \Theta_4 \right)
\right] e^{-2i\omega t} \right\} , \nonumber \\ \label{shgw}
\end{eqnarray}
where $\Theta_1 = ( \left| E_x \right|^2-\left| E_y \right|^2)$,
$\Theta_2 = (E_xE_y^*+E_yE_x^*)$, $\Theta_4 = (E_x^2 - E_y^2)$ and $\Theta_5 = 2E_xE_y$.
For a degenerate electron gas, $\epsilon_F \gg k_BT$,
the coefficients $Q$, $S_1$, $S_2$ are given by
\begin{eqnarray*}
Q &=& - \frac{g e^3 v_g}{8 \pi \hbar^2} \left( \frac{1}{\Upsilon^{1,-1}} + \frac{1}{\Upsilon^{-1,-1}}\right) \\
&& \qquad \times
\left[ \frac{2 C_{-2}}{\Upsilon^{0,1}\Upsilon^{0,-2}}  + p  \left( \frac{C_{-2} v_g}{\Upsilon^{0,1}\Upsilon^{0,-2}} \right)^{\prime}\right]  , \\
&&  + \frac{g e^3 C_{-1} v_g}{8 \pi \hbar^2}
\left( \frac{1}{\Upsilon^{1,2}\Upsilon^{1,-1}} + \frac{1}{\Upsilon^{-1,2}\Upsilon^{-1,-1}}\right)\\
&& \qquad \times
\left[ \frac{1}{\Upsilon^{0,1}} - p \left( \frac{v_g}{\Upsilon^{0,1}} \right)^{\prime} \right]  ,
\end{eqnarray*}
\begin{eqnarray*}
S_1 &=& - \frac{g e^3 v_g}{16 \pi \hbar^2}
\Bigg[ \frac{2C_{-2}}{\Upsilon^{1,1}\Upsilon^{2,-1}\Upsilon^{2,2}} + \frac{2C_{-2}}{\Upsilon^{1,-1}\Upsilon^{2,1}\Upsilon^{2,-2}} \\
&& + \frac{p}{\Upsilon^{1,1}} \left( \frac{v_gC_{-2}}{\Upsilon^{2,-1}\Upsilon^{2,2}} \right)^{\prime}
+ \frac{p}{\Upsilon^{1,-1}} \left( \frac{v_gC_{-2}}{\Upsilon^{2,1}\Upsilon^{2,-2}} \right)^{\prime} \\
&& - \frac{C_{-1}}{\Upsilon^{1,1}\Upsilon^{1,-2}\Upsilon^{2,-1}} - \frac{C_{-1}}{\Upsilon^{1,-1}\Upsilon^{1,2}\Upsilon^{2,1}} \\
&& + \frac{C_{-1}p}{\Upsilon^{1,1}\Upsilon^{1,-2}} \left( \frac{v_g}{\Upsilon^{2,-1}} \right)^{\prime}
+ \frac{C_{-1}p}{\Upsilon^{1,-1}\Upsilon^{1,2}} \left( \frac{v_g}{\Upsilon^{2,1}} \right)^{\prime} \Bigg] ,
\end{eqnarray*}
\begin{eqnarray*}
S_2 &=& \frac{i g e^3 v_g}{16 \pi \hbar^2}
\Bigg[ \frac{2C_{-2}}{\Upsilon^{1,1}\Upsilon^{2,-1}\Upsilon^{2,2}} - \frac{2C_{-2}}{\Upsilon^{1,-1}\Upsilon^{2,1}\Upsilon^{2,-2}} \\
&& + \frac{p}{\Upsilon^{1,1}} \left( \frac{v_gC_{-2}}{\Upsilon^{2,-1}\Upsilon^{2,2}} \right)^{\prime}
- \frac{p}{\Upsilon^{1,-1}} \left( \frac{v_gC_{-2}}{\Upsilon^{2,1}\Upsilon^{2,-2}} \right)^{\prime} \\
&& - \frac{C_{-1}}{\Upsilon^{1,1}\Upsilon^{1,-2}\Upsilon^{2,-1}} + \frac{C_{-1}}{\Upsilon^{1,-1}\Upsilon^{1,2}\Upsilon^{2,1}} \\
&& + \frac{C_{-1}p}{\Upsilon^{1,1}\Upsilon^{1,-2}} \left( \frac{v_g}{\Upsilon^{2,-1}} \right)^{\prime}
- \frac{C_{-1}p}{\Upsilon^{1,-1}\Upsilon^{1,2}} \left( \frac{v_g}{\Upsilon^{2,1}} \right)^{\prime} \Bigg] ,
\end{eqnarray*}
where $(\ldots)^{\prime} \equiv \partial (\ldots) / \partial \epsilon$, $g = 2$ for spin and
all parameters are evaluated on the Fermi surface.
Assuming isotropic, short-range impurities
and scattering times $\tau$, $\tau_2$ independent of energy (as in Section~\ref{s:rat}),
the coefficients $Q$, $S_1$, $S_2$ simplify as
\begin{eqnarray*}
Q^{(\mathrm{s,a})} \!\!&=&\!\! \xi r^{(\mathrm{s,a})} \frac{e^3 v_3}{4\pi \hbar^2 \tau} \frac{1}{\Upsilon^{0,1}\Upsilon^{0,-2}}
\!\left( \!\frac{1}{\Upsilon^{1,-1}} + \frac{1}{\Upsilon^{-1,-1}}\!\right) \! , \label{Qs} \\
S_1^{(\mathrm{s,a})} \!\!&=&\!\! \xi r^{(\mathrm{s,a})}\frac{e^3 v_3}{8\pi \hbar^2 \tau}
\!\left( \!\frac{1}{\Upsilon^{1,1}\Upsilon^{2,-1}\Upsilon^{2,2}}
+ \frac{1}{\Upsilon^{1,-1}\Upsilon^{2,1}\Upsilon^{2,-2}} \!\right) \! , \\
S_2^{(\mathrm{s,a})} \!\!&=&\!\! - \xi r^{(\mathrm{s,a})}\frac{i e^3 v_3}{8\pi \hbar^2 \tau}
\!\left( \!\frac{1}{\Upsilon^{1,1}\Upsilon^{2,-1}\Upsilon^{2,2}}
- \frac{1}{\Upsilon^{1,-1}\Upsilon^{2,1}\Upsilon^{2,-2}} \!\right) \! ,
\end{eqnarray*}
where factor $r^{(\mathrm{s})} = 1$ for symmetric disorder
and $r^{(\mathrm{a})} = -s\zeta \Delta / \gamma_1$ for asymmetric disorder.

The form of the valley currents Eqs.~(\ref{dcw},\ref{shgw}) is consistent with threefold
rotational symmetry (rotations around the $z$ axis by $2\pi/3$).
Terms $\mathrm{Re} \:\! Q$ and $S_1$ are
even functions of $\mathbf{\hat{n}}_z \cdot \mathbf{B}_{\perp}$,
whereas $\mathrm{Im} \;\! Q$ and $S_2$
are odd functions, thus they are zero for $\mathbf{B}_{\perp}=0$.
The dc valley current is only non-zero for incoming linearly-polarized light Eq.~(\ref{lin}),
for which the valley currents may be written as
\begin{eqnarray*}
\mathbf{J}^{(0)} &=& \tfrac{1}{4} E_0^2 |Q| \left[
\boldsymbol{\hat{\textbf{\i}}} \cos (2\theta + \Pi_Q ) - \boldsymbol{\hat{\textbf{\j}}} \sin (2\theta + \Pi_Q ) \right] \\
\mathbf{J}^{(2)} &=& \tfrac{1}{2} E_0^2 |S_1| \cos \left( 2 \omega t - \Pi_1 \right)
\left( \boldsymbol{\hat{\textbf{\i}}} \cos 2\theta - \boldsymbol{\hat{\textbf{\j}}} \sin 2\theta \right) , \\
&& + \tfrac{1}{2} E_0^2 |S_2| \cos \left( 2 \omega t - \Pi_2 \right)
\left( \boldsymbol{\hat{\textbf{\i}}} \sin 2\theta + \boldsymbol{\hat{\textbf{\j}}} \cos 2\theta \right) ,
\end{eqnarray*}
where $\Pi_Q = \mathrm{arg} (Q)$, $\Pi_1 = \mathrm{arg} (S_1)$ and $\Pi_2 = \mathrm{arg} (S_2)$.
This indicates that the incoming polarization angle $\theta$ dictates
the direction of outgoing dc current and the polarization of the second harmonic.
For incoming circularly-polarized light Eq.~(\ref{cir}), the dc current is zero.
For the second harmonic
\begin{eqnarray*}
\mathbf{J}^{(2)} &=& E_0^2 |S_1|
\Big\{
\boldsymbol{\hat{\textbf{\i}}} \cos \left(2 \omega t - \Pi_1 \right)  \\
&& \qquad \qquad - \mu \boldsymbol{\hat{\textbf{\j}}} \sin \left(2 \omega t - \Pi_1 \right) \Big\} ,  \\
&& + E_0^2 |S_2|
\Big\{
\boldsymbol{\hat{\textbf{\i}}} \cos \left(2 \omega t - \Pi_2 - \mu \pi / 2\right)  \\
&& \qquad \qquad - \mu \boldsymbol{\hat{\textbf{\j}}} \sin \left(2 \omega t - \Pi_2 - \mu \pi / 2\right) \Big\} .
\end{eqnarray*}
Thus, the generated current is also circularly polarized, but the sense of the polarization
is reversed as compared to the incoming light.

The valley currents display cyclotron resonances at the same frequencies
as the magnetic ratchet: $\omega_c = \omega$ for the dc current and $\omega_c = \omega$ and $\omega_c = 2\omega$
for the second harmonic.
The valley current is much larger than the magnetic ratchet current per valley,
typically by a factor of $mv_3/[|\mathbf{B}|ed(\gamma_4/\gamma_0 + \delta/\gamma_1)] \sim 1500$
for $|\mathbf{B}| = 1\,$T, although,
obviously, the total current is zero when summed over both valleys.
Although we have considered only linear powers of $\gamma_3$, a similar qualitative picture holds for higher powers.
Even powers of $\gamma_3$ would not appear with the valley index, but they are only capable of coupling
harmonics with values of the angular index $j$ that differ by an even number. Hence, they do not
produce second-order-in-$\mathbf{E}$ currents. Odd powers of $\gamma_3$ can couple harmonics with values
of the angular index $j$ that differ by an odd number [as in Eq.~(\ref{wcouple})],
thus yielding second-order-in-$\mathbf{E}$ currents,
but they always appear with the valley index, giving a total current of zero.

\section{Conclusions}

As detailed in Section~\ref{s:be}, we have determined the dc current, Eq.~(\ref{j0full}),
and the second harmonic generation, Eq.~(\ref{j2full}), for the magnetic ratchet in the semiclassical
regime ($\epsilon_F \gg \hbar \omega$) in a two-dimensional electron system.
These results apply to systems with arbitrary, isotropic dispersion
and energy-dependent scattering rates. For the particular case of bilayer graphene,
we assume a perfectly quadratic dispersion relation $\epsilon = p^2/2m$
and relaxation rates that are independent of energy to produce simplified expressions
for the dc current, Eqs.~(\ref{j0intro},\ref{ms},\ref{ma}),
and second harmonic generation, Eqs.~(\ref{j2intro},\ref{n1f},\ref{n2f}).
We take into account inversion symmetry breaking by disorder and by interlayer asymmetry,
the latter may potentially by induced using an external gate and is thus tunable.
In the presence of a tilted field, we find that the dc current has a resonance
at $\omega_c = \omega$ but that the current value is actually largest at $\omega_c = 0$.
For the second harmonic, however, resonances
at $\omega_c = \omega$ and $\omega_c = 2\omega$ generally produce currents significantly
greater than that at $\omega_c = 0$.


\begin{acknowledgments}
We thank S.~D.~Ganichev and S.~Slizovskiy for useful discussions.
This work was funded by the EU Flagship Project, the ERC Synergy Grant Hetero2D
and EPSRC grant EP/N010345/1.
\end{acknowledgments}


\begin{thebibliography}{99}
\bibitem{notebud}
Coefficients $M_1$, $M_2$, $M_3$ in Eq.~(\ref{j0full}) are related to
$M_0$, $M_L$, $M_C$ in \cite{bud16} by $M_1 = M_L$, $M_2 = M_0$,
$M_3 = i M_C$. These choices are made to be consistent with earlier works
\cite{tara11,khe16} and to ensure the coefficients are real for
$\omega_c = 0$.
\end{thebibliography}
\end{document}